\newif\ifShowKeys
\ShowKeystrue

\documentclass[11pt]{article}
\pdfoutput=1
\usepackage[usenames,dvipsnames]{xcolor}
\usepackage[setpagesize=false,pagebackref=false, linktocpage, bookmarksopen=true, colorlinks=true, linkcolor=Maroon,citecolor=Maroon,urlcolor=Maroon]{hyperref}

 \usepackage[parsep]{collref}

\usepackage{tocloft}

\usepackage{amsmath, amssymb,amsthm}
\usepackage{amsthm}
\numberwithin{equation}{section}

\usepackage{tabto}
\usepackage{bm,environ,mathrsfs,array,arydshln}
\usepackage{booktabs,float,slashed,hyperref}
\usepackage[mathcal]{euscript}
\usepackage{tensor} 						
\usepackage{mathabx}
\usepackage{simpler-wick}
\usepackage[vcentermath]{youngtab}

\usepackage{aurical}
\usepackage[T1]{fontenc}
\usepackage[nodayofweek]{date time}
\usepackage{graphicx,epsfig,epic}
\usepackage{tikz} 
\usetikzlibrary{arrows,decorations.pathreplacing,decorations.markings,snakes}
\tikzset{middlearrow/.style={decoration={markings, mark= at position 0.5 with {\arrow{#1}} ,
}, postaction={decorate}}}
\tikzset{decoration={snake,amplitude=.4mm,segment length=2mm,
                       post length=0mm,pre length=0mm}}

\usepackage{framed}						
\definecolor{shadecolor}{rgb}{0.9996078, 0.984314, 0.960784}

\usepackage{comment}

\allowdisplaybreaks

\definecolor{myred}{RGB}{233, 33, 45}

\newcommand{\bs}{\begin{shaded}}
\newcommand{\es}{\end{shaded}\noindent}
\def\ba#1\ea{\begin{align}#1\end{align}}		
\newcommand{\be}{\begin{equation}}
\newcommand{\ee}{\end{equation}}
\newcommand{\mc}{\mathcal }

\newcommand{\la}{\label}

\newcommand{\lp}{\notag \\ & }

\DeclareMathOperator{\sign}{\text{sign}}
\DeclareMathOperator{\tr}{\text{tr}}

\newcommand{\wt}{\widetilde}

\newcommand{\cf}{\textit{cf.} }
\newcommand{\ie}{\textit{i.e.} }

\makeatletter
\DeclareFontFamily{OMX}{MnSymbolE}{}
\DeclareSymbolFont{MnLargeSymbols}{OMX}{MnSymbolE}{m}{n}
\SetSymbolFont{MnLargeSymbols}{bold}{OMX}{MnSymbolE}{b}{n}
\DeclareFontShape{OMX}{MnSymbolE}{m}{n}{
<-6>  MnSymbolE5
   <6-7>  MnSymbolE6
   <7-8>  MnSymbolE7
   <8-9>  MnSymbolE8
   <9-10> MnSymbolE9
  <10-12> MnSymbolE10
  <12->   MnSymbolE12
}{}
\DeclareFontShape{OMX}{MnSymbolE}{b}{n}{
<-6>  MnSymbolE-Bold5
   <6-7>  MnSymbolE-Bold6
   <7-8>  MnSymbolE-Bold7
   <8-9>  MnSymbolE-Bold8
   <9-10> MnSymbolE-Bold9
  <10-12> MnSymbolE-Bold10
  <12->   MnSymbolE-Bold12
}{}

\let\llangle\@undefined
\let\rrangle\@undefined
\DeclareMathDelimiter{\llangle}{\mathopen}%
 {MnLargeSymbols}{'164}{MnLargeSymbols}{'164}
\DeclareMathDelimiter{\rrangle}{\mathclose}%
 {MnLargeSymbols}{'171}{MnLargeSymbols}{'171}
\makeatother

\catcode`,\active

\catcode`\,12

\def\XXint#1#2#3{{\setbox0=\hbox{$#1{#2#3}{\int}$}
     \vcenter{\hbox{$#2#3$}}\kern-.5\wd0}}

\renewcommand{\l}{\lambda}

\newcommand{\Z}{\mathbb{Z}}

\newcommand{\bw}{\bar w}

\newcommand{\CP}{{\mathbb{C}{\rm P}}}

\newcommand{\gs}{g_{\rm s}}

\renewcommand{\l}{\lambda}

\newcommand{\rf}[1]{(\ref{#1})}

\newcommand{\Ntot}{\nu}
\newcommand{\OO}{\mathsf{O}}

\newcommand{\s}{\sigma}

\def \tc {{_{\rm I}}}  \def \ttc {{_{\rm II}}}

\def \ha {\tfrac{1}{2}}
\def \foot {\footnote}
\def \T {{\rm T}}  
 
\def \iffa {\iffalse}
  \def \ha {\frac{1}{2}}
  
\def \adsz  {AdS$_{4}\times S^{7}/\mathbb{Z}_{k}$ }
\def \adszz  {AdS$_{4}\times S^{7}/\mathbb{Z}_{k}$}

\def \vol {{\rm vol}}

\def \te {\textstyle} \def \ZZ {{\mathbb Z}}
\def \l {\lambda}  \def \ov {\over}

\def \no {\nonumber}

\def \ZZ {{\mathbb Z}}
\def \td {\tilde}

\def \RL   {R} 
\def \ov {\over}
\def \ci {\cite}
\def \sfm {{\rm w}}
\def \SS  {{\rm S}}
\def \IS    {{S}}
\def \RP { {\mathbb{R}{\rm P} }   }

\def \ed {
\bibliography{BT-Biblio}
\bibliographystyle{JHEP-v2.9}
\end{document}
}

\def \edd {\end{document}
}

\def \adsz  {AdS$_{4}\times S^{7}/\mathbb{Z}_{k}$ }

\def \inst {{\rm inst}}
\def \del {\partial}
\def \lpl  {\ell_{\rm P}}
\def \Z  {{\rm Z}} 
\def \del {\partial}

 \def \z  {\zeta}\def \s  {z}
 
 \def  \Ntot  {{\gamma}}
\def \third {{1\ov 3}}
  \def \Q  {{\cal Q}}

\begin{document}


\begin{titlepage}
\begin{tabbing}
\date{\today}
\hspace*{11.5cm} \=  \kill 
\>  PUPT-2645\\
\>  Imperial-TP-AT-2023-04 \\
\> 
\end{tabbing}

\vspace*{15mm}
\begin{center}

\centerline{\large\sc Instanton contributions to the ABJM free energy}\vskip 4pt
\centerline{\large\sc from quantum M2 branes}


\vspace*{10mm}

{\large M. Beccaria${}^{\,a}$, S. Giombi$^{\,b}$, A.A. Tseytlin$^{\,c, }$\footnote{Also on  leave from  the Institute for Theoretical and Mathematical Physics (ITMP) of MSU    and Lebedev Institute.}} 

\vspace*{4mm}
	
${}^a$ Universit\`a del Salento, Dipartimento di Matematica e Fisica \textit{Ennio De Giorgi},\\ 
		and I.N.F.N. - sezione di Lecce, Via Arnesano, I-73100 Lecce, Italy
			\vskip 0.1cm
${}^b$  Department of Physics, Princeton University, Princeton, NJ 08544, USA
			\vskip 0.1cm
${}^c$ Blackett Laboratory, Imperial College London SW7 2AZ, U.K.
			\vskip 0.1cm
\vskip 0.2cm {\small E-mail: \texttt{matteo.beccaria@le.infn.it},\ \texttt{sgiombi@princeton.edu}, \ \texttt{tseytlin@imperial.ac.uk}}
\vspace*{0.8cm}
\end{center}

\begin{abstract}  
	{We present a quantum M2  brane   computation of the 
 instanton prefactor in the  leading non-perturbative  contribution to the 
 ABJM  3-sphere free energy  at large $N$ and fixed level $k$. Using supersymmetric localization, such instanton contribution was found earlier to take the form
 $F^{\rm inst}(N,k)  =   - ({\sin^{2} \frac{2\pi}{k}} )^{-1} \, {\rm exp} ({-2\pi  \sqrt{2N\ov k }}) + ... $. 
 The exponent   comes  from the action of an M2 brane  instanton wrapped on $S^3/\ZZ_k$, 
 which represents the M-theory  uplift of the  $\CP^1$ instanton  in  type IIA string theory on AdS$_4 \times \CP^3$. 
 The IIA  string  computation of the  leading large $k$ 
 term in the instanton prefactor 
was  recently performed   in   arXiv:2304.12340.
Here   we find    that the  exact   value of the  prefactor $({\sin^{2} \frac{2\pi}{k}})^{-1} $ 
  is reproduced  by the 1-loop term in the M2  brane partition  function expanded near the $S^3/\ZZ_k$ instanton configuration.  As in the  Wilson loop example   in   arXiv:2303.15207, 
   the quantum M2
    brane  computation 
   is well defined   and  produces  a finite result  in exact agreement with localization.}
\end{abstract}	
	
	\iffa
	
We present a quantum M2  brane   computation of the 
 instanton prefactor in the  leading non-perturbative  contribution to the 
 ABJM  3-sphere free energy  at large $N$ and fixed level $k$. Using supersymmetric localization, such instanton contribution was found earlier to take the form
 $F^{\rm inst}(N,k)  =   - ({\sin^{2} \frac{2\pi}{k}} )^{-1} \, {\rm exp} ({-2\pi  \sqrt{2N\ov k }}) + ... $. 
 The exponent   comes  from the action of an M2 brane  instanton wrapped on $S^3/\ZZ_k$
 which represents the M-theory  uplift of the  $\CP^1$ instanton  in  type IIA string theory on AdS$_4 \times \CP^3$. 
 The IIA  string  computation of the  leading large $k$   term in the instanton prefactor 
was  recently performed   in   arXiv:2304.12340. Here   we find    that the  exact  value of the    
prefactor $({\sin^{2} \frac{2\pi}{k}})^{-1} $  is reproduced  by the 1-loop term in the M2  brane partition  function expanded near the $S^3/\ZZ_k$ instanton configuration.  As in the  Wilson loop example   in   arXiv:2303.15207, 
 the quantum M2  brane  computation   is well defined   and  produces  a finite result  in exact agreement with localization.
		
	\fi

\end{titlepage}
\tableofcontents
\vspace{1cm}

\parskip 0.1cm

\section{Introduction}

The  ABJM duality   \ci{Aharony:2008ug}  between  
the  supersymmetric Chern-Simons-matter 
 theory and  11d M-theory on  \adszz, combined with exact localization results \ci{Pestun:2016zxk}, 
 provides a possibility to shed  light on the structure of 
 M-theory as a theory of quantum supermembranes. 
 
A recent  remarkable example  was  provided  in \ci{Giombi:2023vzu}, which considered 
the  $1\ov 2$-BPS circular Wilson loop expectation value  $\langle W_\ha \rangle 
$  in the $U(N)_k \times U(N)_{-k}$ ABJM theory at  large $N$    and fixed level $k$. This
 has a dual description in terms of  an M2 brane  wrapped   on AdS$_2 \times S^1 $ \ci{Sakaguchi:2010dg}
   in the M-theory background AdS$_4 \times S^7/\mathbb Z_k$. 
 The localization result \ci{Klemm:2012ii} \ \ 
 $\langle W_{\frac{1}{2}}\rangle =   \frac{1}{2\, \sin(\frac{2\pi}{k})} \, e^{ \pi  \sqrt{2N\ov k }\,  }  + ...
  $   has  the  exponential factor  that   comes  from the   classical value of the M2 brane  action,  
  while the 
$k$-dependent prefactor  was  exactly reproduced \ci{Giombi:2023vzu}
  by the one-loop term in the partition function of the  quantum  M2 brane. 
  This  consistent quantum M2 brane computation naturally suggests extensions to M-theory calculations for other observables 
  that may be similarly compared to localization results. 
 
 Indeed, here we  present an analogous quantum M2  brane   computation of the 
 instanton prefactor in the  localization result for the  leading large $N$   non-perturbative   contribution to the 
 ABJM   free energy $F$  on  the 3-sphere, which has the form \ci{Drukker:2011zy,Hatsuda:2013gj} \ \ 
 $F^{\rm inst}(N,k)  =   - \frac{1}{\sin^{2} (\frac{2\pi}{k})} \, e^{-2\pi  \sqrt{2N\ov k }}+ ... $. 
 Here the exponent   comes  from the action of an M2 brane  instanton with $S^3/\ZZ_k$ world-volume geometry. Such M2 instanton wraps the 11d
 circle $S^1$ and a $\CP^1$ in $\CP^3$, and 
 it represents the M-theory uplift of the    $\CP^1$ instanton  in  type IIA string theory on AdS$_4 \times \CP^3$ 
 \ci{Cagnazzo:2009zh}.  The IIA superstring   computation of the  leading large $k$ 
 term in this instanton prefactor $\frac{1}{\sin^{2} (\frac{2\pi}{k})} \to {k^2\ov (4 \pi)^2}  = { 2 T \ov \pi \gs^2} $ 
 ($T$  is the string tension  and $\gs$ is the string coupling)  was  recently presented in a remarkable 
  paper \ci{Gautason:2023igo}.  Here   we show   that the  exact    prefactor $\frac{1}{\sin^{2} (\frac{2\pi}{k})} $ 
  is reproduced  by the 1-loop term in the M2    brane partition  function expanded near the $S^3/\ZZ_k$ instanton configuration.  As in the  Wilson loop example   \ci{Giombi:2023vzu}, the quantum M2
    brane  computation 
   is well defined   and  produces  a finite result which is  in exact agreement with the localization prediction. 
 
We shall start in section 2   with a review of the relevant 
 localization results for the  large $N$ non-perturbative  contributions 
  to the ABJM free energy $F$ on 3-sphere. 
 We shall  note a surprising similarity  between the instanton prefactor
  in the non-perturbative part of  
 $F$  and in the leading 
 perturbative   term in the expectation  value of the Wilson loop. 
 
 In section 3  we  shall  argue that  the  gauge theory  free energy   should be matched to  the ``first-quantized'' 
 M2 brane partition function.   The leading large $N$  perturbative 
 terms  should be captured by the 11d supergravity action  plus  higher derivative  corrections 
 (see \ci{Beccaria:2023hhi}  and refs. there),
  while
  non-perturbative contributions    should  come from  M2 brane instanton
  contributions.

 Section 4 will be devoted to the  classical  solution for the M2 brane  wrapped on $S^3/\ZZ_k \subset S^7/\ZZ_k$ 
 and the  Lagrangian for the  bosonic and fermionic quadratic fluctuations   around  it. 
 We will use a static gauge,  where the 11d circle is  identified with one periodic   M2 coordinate. 
Expanding in Fourier modes on this circle, we  may  represent  the M2 brane 3d fluctuation action 
 as   an  action   for an infinite Kaluza-Klein tower (labeled  by $n=0, \pm 1, \pm 2, ...$)  of 2d 
  fields on $\CP^1$, in the presence of a background $U(1)$ gauge field of a magnetic monopole (originating from the Hopf fibration representation of $S^3/\ZZ_k$). The lowest $n=0$ level  corresponds to  the type IIA  string 
 fluctuations near the $\CP^1$  instanton  already discussed   in   \ci{Cagnazzo:2009zh,Gautason:2023igo}. 
 
 The determinants  of the resulting   bosonic and fermionic operators on $S^2$  for  charged massive 2d fields in 
 the magnetic monopole background are computed in section 5. 
   
 In section  6 we    perform the sum over the level 
  $n$ of all the    2d  fluctuation contributions. We  observe that the
 final  result   for the M2 brane 1-loop correction 
 is UV finite provided  one uses the standard  analytic (Riemann $\zeta$-function) regularization, in agreement with expectation of  no  1-loop log UV divergences in a  3d theory. 
 We also discover that  
 almost  all   finite  bosonic  and fermionic terms  mutually cancel  and  
 the  1-loop M2 brane partition function ${\rm Z}_1$ appears to effectively ``localize'' to the contribution of 
  just  two  bosonic 1d degrees of freedom on $S^1$, i.e.
   $\log {\rm Z}_1\to   - \log {\rm det}' \big({\te -  {k^2\ov 4} {d^2\ov d s^2}  }-1\big)
=- 2 \sum_{n=1}^{\infty}\log\big(\tfrac{k^{2}}{4}n^2 -1\big)$.  
Regularizing this  1d determinant  in the standard way  we get 
the  $  \frac{1}{ \sin^2(\frac{2\pi}{k})}$ prefactor   matching  the 
 localization  result (up to   an integer factor that requires consideration of  string-level  0-modes  as in \ci{Gautason:2023igo}). 

We conclude  in section 7  by making some  remarks on several  open problems. 
In particular, we    comment on the role of the 
$\RP^3 \subset \CP^3$   M2  brane  instanton (or the D2 brane instanton  \ci{Drukker:2011zy}  in the type IIA string limit) 
 with  some  details about  this second type of instanton contributions to the  free energy  provided 
   in Appendix A, where we also comment on the  cases of $k=1,2$  which need  special consideration. 

\section{Free energy  from localization}

Let us start with a review of  the localization result  for the partition function 
$Z(N,k) $  of the  $U(N)_k \times U(N)_{-k}$  ABJM   theory 
on $S^3$. 
We shall assume that  $k >2$  (and in general finite). 
As a function of $N$   the partition function    can be  represented as a sum of a 
perturbative  part (given by a series in $1\ov \sqrt N $)  and a non-perturbative part 
involving  factors like $e^{- h(k)\sqrt N}$  that are exponentially  suppressed at large $N$, i.e.
\be 
Z= Z^{\rm p}(N,k) + Z^{\rm np}(N,k)  \ . \la{x}\ee
In the Fermi gas approach\cite{Marino:2011eh} the  localization expression
 for  $Z(N,k) $  is expressed\footnote{The localization matrix model representation for the partition function of the ABJM theory on $S^{3}$ was first derived in \cite{Kapustin:2009kz} and later mapped to a lens space matrix model solvable in planar limit  \cite{Marino:2009jd}. 
Higher order  $1/N$  corrections were computed in \cite{Drukker:2011zy,Drukker:2010nc} and resummed in \cite{Fuji:2011km} neglecting non-perturbative corrections.
}
   in terms of the grand potential $J(\mu, k)$ 
of a non-trivial fermionic system as\footnote{To be precise, the integration contour in (\ref{2.1}) corresponds to the use of the so-called modified grand potential, see 
 \cite{Hatsuda:2012dt} for details.}
\be
\la{2.1}
Z(N,k) \equiv e^{-F(N,k) }=  \int_{-i\infty}^{i\infty}\frac{d\mu}{2\pi i} \ e^{J(\mu, k)-N\mu}\ .
\ee
The grand potential $J(\mu,k)$  may be   split into 
  the sum of the perturbative (polynomial in $\mu$)  
  and non-perturbative (suppressed at large $\mu$)  parts
\be
J(\mu, k) = J^{\rm p}(\mu,k)+J^{\rm np}(\mu,k) \    \la{2.2}
\ee
where 
\be
J^{\rm p}(\mu, k) = \frac{1}{3}C(k)\, \mu^{3}+B(k)\, \mu+A(k), \qquad C(k) = \frac{2}{\pi^{2}k}, \qquad B(k) = \frac{k}{24}+\frac{1}{3k} \ . 
\ee
Here  $A(k)$ is the so-called constant map contribution
first identified in \cite{Hanada:2012si} and admitting the following  integral representation \cite{Hanada:2012si,Hatsuda:2014vsa}
\be\la{2.4} 
A(k) = -\frac{\zeta(3)}{8\pi^{2}}\,\Big(k^{2}-\frac{16}{k}\Big)+\frac{k^{2}}{\pi^{2}}\,\int_{0}^{\infty}dx\, \frac{x}{e^{kx}-1}\,\log(1-e^{-2x}).
\ee
If (\ref{2.1}) is evaluated keeping only the perturbative part  $J^{\rm p}(\mu, k) $ in \rf{2.2} one finds 
the  perturbative part of  $Z$  which is  expressed in terms of  the Airy  function  
\be\la{2.5}
Z^{\rm p}(N,k) = C(k)^{-\frac{1}{3}}\, e^{A(k)}\, \text{Ai}\big[C(k)^{-\frac{1}{3}}\big(N-B(k)\big)\big] \ .
\ee
As for  the  non-perturbative contributions  to $Z$,  the structure of $J^{\rm np}(\mu, k)$
implies  that 
there are  two basic types  of  exponential  corrections.\footnote{Note that 
in the saddle point evaluation of (\ref{2.1}), the large $\mu$ and the large $N$  limits 
are correlated due to the form of the perturbative grand potential.}
 Written as the exponentially suppressed  contributions $F^{\rm np}$  to the free  energy  
\ba
F\equiv  - \log  Z= F^{\rm p}  + F^{\rm np}  \ , \la{xx} \ea
where the large $N$  expansion of the perturbative part   follows  from \rf{2.5}
(see \cite{Beccaria:2023hhi} for details)
 \be
  F^{\rm p}  =- \log Z^{\rm p}=  \te  \frac{1}{3} \sqrt{2}  \pi  k^{1/2}  N^{3/2}-\frac{\pi}{24\sqrt 2}\big( k^2 +8 \big)\,k^{-1/2}\,N^{1/2}  
 +\frac{1}{4}  
\log  {32  N\ov k}  - A(k)  + \OO ( N^{-1/2}) 
     \ ,  \la{xxx}  \ee
they  are given  by the double sum  \ci{Drukker:2011zy,Hatsuda:2013gj}
\be 
\la{2.6}
F^{\rm np} = \sum^\infty_{n_{\tc}, n_{\ttc}=0} f_{n_{\tc}, n_{\ttc}}(N,k)\,\exp\Big[-2\pi\sqrt{N}\, \Big(n_{\tc} \sqrt {2\ov  k}+n_{\ttc}  \sqrt{k\ov  2}\, \Big)\Big] \ . 
\ee
 In   the type IIA   string theory regime  (i.e. in the limit of large $N$ and $k$ 
 with $\l= {N\ov k}$=fixed) 
   these   may be interpreted as  the  contributions 
     of  the    string world-sheet instantons (wrapping $\CP^1$ in $\CP^3$
 \cite{Cagnazzo:2009zh})    and  of the 
  D2-brane instantons (wrapping a 3-cycle  $\RP^3=S^3/\ZZ_2$ 
  in $\CP^3$) respectively  \ci{Drukker:2011zy}. 
 In  the  M-theory regime  (i.e. for large $N$ with fixed $k$), 
 the world-sheet instantons correspond to  the M2 brane instantons wrapping the 11d circle and a $\CP^1$ in $\CP^3$, 
  i.e.   $S^3/\mathbb{Z}_k\subset S^7/\ZZ_k$, 
   while the D2 instantons 
   correspond to the M2 instantons wrapping 
   the  analog of $\RP^3\subset \CP^3$  3-cycle in   
   $S^7/\ZZ_k$.
 
  We denote the numbers of the two kinds of instanton respectively as  $n_{\tc}$ and $n_{\ttc}$. 
 The   factors in the exponents in (\ref{2.6}) correspond to the classical volumes 
 of the two  types of  the membrane instantons. 
The  terms with both  $n_{\tc}$ and $n_{\ttc}$  nonzero can be thought of as  ``bound state'' contributions \cite{Hatsuda:2013gj}.

For $k >2$  the  dominant  non-perturbative contribution to  \rf{2.6} 
  comes   from the  $S^3/\mathbb{Z}_k$
 instanton, i.e.  from the 
 $n_{\tc}=1$, $n_{\ttc}=0$  term in the sum. 
 Below we shall focus on this  leading term  and 
 simply refer to it as the  ``instanton'' contribution.

The  $n_{\ttc}=0$  part of \rf{2.6} originates  from the following contribution to 
  $J^{\rm np}(\mu, k)$    \cite{Hatsuda:2012dt}  
\be\la{y}
J^{\rm np} (\mu, k) = \sum_{n_{\tc}=1}^{\infty}d_{n_{\tc}}(k)\, e^{-\frac{4n_{\tc}}{k}\mu } + ...\ ,\ \ \ \ 
\ee
where the   function $d_{n_{\tc}}(k)$ may be determined using that the ABJM matrix integral is dual to the partition function
of a  topological string theory on $\mathbb{P}^{1}\times \mathbb{P}^{1}$. 
In particular,  the   $n_{\tc}=1$  instanton term  has  the coefficient  \cite{Hatsuda:2012dt} 
\be\la{yy}
d_1(k)  = \frac{1}{\sin^{2} (\frac{2\pi}{k})} \ . 
\ee
From \rf{2.1}   we  find  that   the  contribution of the 
1-instanton term to  the non-perturbative part of the partition function 
$Z^{\rm np}(N,k)$  can be expressed in terms of the perturbative  part $Z^{\rm p}$   in \rf{2.5} as 
\ba
Z^{\rm inst}(N,k) &= \int_{-i\infty}^{i\infty}\frac{d\mu}{2\pi i} \ e^{J^{\rm p}(\mu,k)-N\mu}\, d_1(k)\, e^{-\frac{4}{k}\mu} = d_1(k) \, Z^{\rm p}(N+\tfrac{4}{k},k)\ . \la{yyy}
\ea
The corresponding 1-instanton  term  in  the non-perturbative  part of  free energy  \rf{xx}  is then (cf. \rf{2.5},\rf{2.6}) 
\ba
\la{213}
F^{\rm np}(N,k) &= F^{\rm inst}(N,k)+ \cdots, \\
\la{2.11}
F^{\rm inst}(N,k) & =  - d_1(k) \frac{\text{Ai}[C(k)^{-\frac{1}{3}}(N-B(k)+\tfrac{4}{k})]}{\text{Ai}[C(k)^{-\frac{1}{3}}(N-B(k))]}\no\\
&   =  F^{\rm inst}_1(N,k) \, \Big[1  +   {\te   {\pi \ov  \sqrt {2k}  }  {k^2-40\ov 12k } {1 \ov \sqrt N} }+ ...       \Big] \ , \\
 F^{\rm inst}_1(N,k)  &= - d_1(k) \,  e^{-2\pi\sqrt{N} \sqrt {2\ov  k} \,  } =
 - \frac{1}{\sin^{2} (\frac{2\pi}{k})} \, e^{-2\pi  \sqrt{2N\ov k }\,  } \ . \la{z}
\ea
Here $F^{\rm inst}_1$  is the leading   large $N$  term in the 1-instanton 
contribution $F^{\rm inst}(N,k)$. 

It is  interesting to notice  a  close resemblance of the expression for $F^{\rm inst}(N,k) $ 
in \rf{2.11}   and the one for the perturbative part of  
  the expectation value 
of the $1\ov 2$-BPS circular Wilson loop 
  in the  ABJM theory  
  derived  using   localization in \cite{Klemm:2012ii} 
  (see  also \cite{Drukker:2010nc,Herzog:2010hf,Fuji:2011km,Marino:2011eh})\foot{We use the same  normalization  of $\langle W_{\frac{1}{2}}\rangle$ as 
in \ci{Giombi:2020mhz,Giombi:2023vzu}.}  
\ba
\langle W_{\frac{1}{2}}\rangle =& \frac{1}{2\, \sin(\frac{2\pi}{k})} \frac{{\rm Ai}[C(k)^{-\frac{1}{3}}(N- B(k) - {2\ov k} )]}{{\rm Ai}[C(k)^{-\frac{1}{3}}
(N-  B(k) )]}\no \\
=&   \frac{1}{2\, \sin(\frac{2\pi}{k})} e^{ \pi  \sqrt{2N\ov k }\,  } \Big[ 1  - 
   {\te   {\pi \ov  \sqrt {2k}  }  {k^2+32\ov 24k } {1 \ov \sqrt N} }+ ...       \Big]  
  \ .
\label{zz}
\ea
Compared to \rf{2.11}  here  the  prefactor  is $ \frac{1}{2\, \sin(\frac{2\pi}{k})}$ 
instead of   $- \frac{1}{ \sin^2(\frac{2\pi}{k})}$  in \rf{z} and  the argument 
of  the Airy  function  in the numerator is 
$N-B(k) -  {2\ov k} = N-\frac{k}{24}-\frac{7}{3k}$ instead of 
$N-B(k) +\frac{4}{k}= N-\frac{k}{24}+\frac{11}{3k}$. 
These  different shifts explain   why 
   the  coefficients  in the  exponentials  in \rf{z} and \rf{zz}  
differ by factor of $-2$.\foot{Indeed,  the large  $x$ asymptotics of 
${\rm Ai}(x) \sim    x^{-1/4}  e^{ - 2/3  x^{3/2}} $   implies that 
for the ratio with the arguments $N-B(k) +  {a\ov k}$ and $N-B(k)$ 
the  asymptotics is $e^{ - a \pi \sqrt{  N \ov 2k}}$.}

This factor of $-2$   has a  string theory (or   wrapped 
M2 brane) interpretation:   the  regularized 
area  of the  AdS$_2$  minimal   surface  in the case  of the Wilson loop 
is $-2 \pi$  (minus area of a disk), while the area   of  $\CP^1$  is $+4 \pi$.\foot{In more detail, for the WL we have an AdS$_2$  inside AdS$_4$ of radius  $\RL/2$ 
 so the area  is $-\ha \pi  \RL^2$. 
 For the M2 brane  instanton, the  $\CP^3$
  factor in the metric has radius  $\RL$  but the   $\CP^1$
   we wrap around  is a sphere of  radius $\RL/2$, so
   the resulting area   is $\pi  \RL^2$. 
   Thus the  relative factor is  still  $-2$   as  above. 
}
Also,   the fact that  the prefactor in
the instanton contribution to the free   energy in 
\rf{z}   is proportional to the square of the prefactor   in the Wilson loop in \rf{zz}
 may be  attributed to the 
a heuristic expectation that  the  partition   function on a  2-sphere 
is related to a square of the partition function on a disk.

Our aim below will be to reproduce \rf{z} on the dual   M-theory side 
  by a quantum M2  brane computation, 
   in full analogy to how that was done in \ci{Giombi:2023vzu}  for the leading  term
     in the Wilson loop  expression in  \rf{zz}.

\section{Free energy from M2 brane partition function}

Motivated by  the expected duality  between the ABJM theory   and M-theory 
on \adsz  it is natural  to expect  that  the perturbative 
part of the   free energy  \rf{xxx}   should be reproduced   by some 
higher derivative extension of the 11d supergravity action   evaluated on the 
 \adsz   background (for a recent discussion and references to related work see 
  \ci{Bobev:2022eus,Beccaria:2023hhi}). Indeed, it was found in 
\ci{Drukker:2010nc} that the leading $N^{3/2}$ term in \rf{xxx} 
is matched by the on-shell  value of the Euclidean
11d supergravity action.\foot{More precisely, if one directly  evaluates  the 
11d action on the \adsz  solution one gets the  leading term in \rf{xxx} 
with an extra factor of $- \ha$ (see  discussion in Appendix B of \ci{Beccaria:2023hhi}).  There  is a subtlety here:   
 as \adsz  is an ``electric'' 11d  solution,  one may use a prescription that  the  sign of the 
 flux $F^2_{mnkl}$  term  is to be reversed  when evaluating the on-shell  value of the 11d action
 (this prescription is  equivalent to adding a particular boundary term; a
 related observation is that  adding a total derivative  4d term
 $ \int  d^4 x\,    \epsilon^{mnkl} F_{mnkl} $ changes the value of the 4d cosmological constant  \ci{Aurilia:1980xj}). 
 This then gives 
   the same   value  as
 found from the   effective 4d action   reconstructed to have the  same   AdS$_4$ space   as its solution
 (see \ci{Kurlyand:2022vzv,Aguilar-Gutierrez:2022kvk}  and refs. there 
  for related discussions).
  It is the  latter  4d action  that was used as a  starting point in \ci{Drukker:2010nc} following 
  \ci{Emparan:1999pm}.}
 
More  generally, we shall 
conjecture that  the   gauge  theory  free energy 
should be  reproduced   by some   properly defined    supermembrane  partition function, 
\be \la{31} 
F \sim   Z_{\rm M2}\,,  \qquad \qquad     Z_{\rm M2}=  \int [d x\,  d\theta ]  \ e^{- \IS_{\rm M2} [x, \theta] } \ ,  
\ee
where   $\IS_{\rm M2}$ is the M2  brane   action on \adsz   with the dimensionless 
coefficient of the  effective tension 
($\RL$  is the radius of $S^7$  or twice the radius of AdS$_4$, see \rf{3.1} below)
\be\la{310}
\T_{2} \equiv \RL^{3}  T_2={1\ov (2\pi)^{2}}   \frac{\RL^{3}}{\ell_{P}^{3}}=
{\sqrt {2k}\ov \pi} \sqrt {N} \ , \qquad\qquad  \frac{\RL}{\ell_{P}}= (32 \pi^2  Nk)^{1/6}  \ .
\ee
Then  for fixed $k$  (or fixed radius of the 11d circle) 
 the semiclassical  large $\T_2$  expansion  of  $ Z_{\rm M2}$  should be equivalent 
 to the large $N$  expansion on the gauge theory side.

 One may further   conjecture   that the   perturbative part of  $ Z_{\rm M2}$ 
 in the  large  $\T_{2}\sim { \sqrt N}$  limit  may be    captured   by   an  expansion near
  ``point-like''  
 M2  branes or, more precisely,  degenerate
     3-surfaces  with a  topology  of $S^1$ times a point  which have zero 3-volume. 
 At the same time,  the    non-perturbative  $e^{-  a \T_2}=  e^{- a{\sqrt {2k}\ov \pi}  \sqrt N}$  
 contributions may come from saddle points  with non-vanishing   3-volumes, e.g.  from 
 M2 branes wrapping the M-theory circle and a $\CP^1 \subset \CP^3$, or a 3-cycle in $\CP^3$ 
  (and their superpositions).
 Symbolically, we may write
 \be \la{u}
 Z_{\rm M2} = Z^{(0)}_{\rm M2} + Z^\inst_{\rm M2} + ... \ .  \ee
 Here the first term (coming from contributions of ``degenerate'' M2 brane surfaces) 
when   expanded at large $k$ 
 should represent the sum of all perturbative tree level  plus  higher loop 
 type IIA  string  corrections to the on-shell  value of the partition function.

 A  motivation   for this suggestion  comes from considering the
 perturbative  type IIA   string limit ($k \sim N \gg 1$)  in which the radius  of the
   11d circle is  small and thus the  membrane partition 
   function should   effectively reduce to the type IIA  string partition function (cf. \ci{Duff:1987bx,Achucarro:1989dd,Berman:2006vg,Uehara:2010xi,Meissner:2022lso}).  The latter, expanded in the  inverse  string  tension, 
    should be  closely related to the low-energy  string 
     effective action \ci{Fradkin:1985ys,Tseytlin:1985kh,Tseytlin:1988tv,Tseytlin:1988rr,Ahmadain:2022tew}
     and thus also to the  on-shell value of the latter.\foot{To see directly 
      how a  non-zero on-shell   value of the string effective action  is   reproduced 
     in the case of AdS space  factors  and RR   backgrounds  
     requires understanding how boundary terms are captured by the string path integral, which is currently an open problem.}
 The  contribution of the type IIA  string instanton  saddle  point  \cite{Cagnazzo:2009zh,Gautason:2023igo}
 will then  naturally supplement   the perturbative part of the string partition function. 
 
One should add of course  a reservation  that,  as  the M2  brane  action is highly non-linear  (even its bosonic  part does not become quadratic  in any gauge),
it is not clear how to define its  expansion near 
a degenerate  ``point-like'' membrane  configuration;  this 
apparently requires a non-perturbative  approach to the corresponding 
 quantum  3d world-volume theory.\foot{In particular,  going off shell,
  it is not even clear how one would  compute  the leading terms  of 
   the 11d supergravity action  by starting with the M2  brane path integral.} 
  
  In contrast, the  semiclassical expansion of the M2  brane path integral near a  classical solution  with a non-zero 3d volume 
  is   well defined 
     \cite{Duff:1987cs,Bergshoeff:1987qx,Mezincescu:1987kj,Forste:1999yj, Drukker:2020swu}
     as  in this case   one is able  to fix a static gauge and thus  develop the   standard 
     gaussian perturbation theory. 
   A remarkable recent 
   example  is the computation \ci{Giombi:2023vzu} of the Wilson loop 
   prefactor in \rf{zz} from the  quantum M2   fluctuation determinants near the corresponding AdS$_2 \times S^1$ minimal 3-surface.

\def \CC {{\mathbb  C}}

Below we will perform  a similar semiclassical  computation in the case of the M2 brane instanton wrapping 
$S^3/\ZZ_k $,  reproducing the $1\ov \sin^2 {2\pi\ov k}$ prefactor in 
\rf{z}  from  the corresponding 1-loop fluctuation determinants.

To be able  to  make a precise  comparison  to the localization   result  for the free energy one is to decide 
about the proportionality  coefficient in \rf{31}. We  will assume    
  (as was also done in the string-theory limit in \ci{Gautason:2023igo}) that
\be \la{sss}
F= -  Z^{(0)}_{\rm M2} -  Z^\inst_{\rm M2} + ...= ( \SS_{\rm sugra}+... )
-  Z^\inst_{\rm M2} + ...   \ .  \ee
This assumption is  based on  the expectation 
 that in the string theory  limit the (on-shell  value of)   string effective action  
should be  given by  minus   
 string partition function as was originally  suggested 
 in  \cite{Tseytlin:1985kh}.\foot{This is true,  e.g.,  
 in the open string theory where the string partition function on the disk
in conformal gauge 
 is negative-definite as the regularized  value of the $SL(2,\mathbb{R})$ M\"obius  volume  is negative, $\vol(SL(2,\mathbb{R}))= \vol({\rm AdS}_2) \, \vol(S^1) =-4\pi^2$
 (cf. \ci{Fradkin:1985qd,Tseytlin:1987ww,Liu:1987nz,Andreev:1988cb,Eberhardt:2021ynh,Mahajan:2021nsd}).
In  the  closed  bosonic  string sigma model on $S^2$ near a  conformal point  one has 
$Z_{\rm str} = {1\ov \Omega} 
  Z,$  where 
 $Z=\int d^D x \sqrt G\,   e^{-2\phi}  {\rm exp } ( C \, 
 \chi \, \log \Lambda + ... )$, \, 
 $ C 
 = {1\ov 6}  D  + ...$, \, 
 $\chi(S^2) =2$  and $\Omega$ is the regularized   volume of the $SL(2,\CC)$ M\"obius group, 
$\Omega =  \vol(SL(2,\CC)) = \vol({SO(1,3)\ov SO(3)}) \, \vol(SO(3)) = - 4 \pi^3\log \Lambda$, where $\vol({SO(1,3)\ov SO(3)})= \vol({\rm AdS}_3) = -2 \pi \log \Lambda $. Then    $S_{\rm str}= - Z_{\rm str}= 
  {1\ov 2 \pi^3}    \int d^D x \sqrt G\,  e^{-2\phi} \, C $.   
  In general, one is to replace ${1\ov \log \Lambda}$  by ${d\ov d \log \Lambda}$   (and renomalize the fields) \ci{Tseytlin:1988tv,Tseytlin:1988rr,Tseytlin:2006ak}
  getting the effective action  $S_{\rm str}= 
  {1\ov 2 \pi^3}    \int d^D x \sqrt G\,   e^{-2\phi}  \, \td \beta^\phi, \ \  \td \beta^\phi= C + ... 
   =   {1\ov 6}  (D-26)  - {1\ov 4} \alpha' R^{(D)}+ ... $,   which thus     has the  right sign  for   a Euclidean action. 
 }

This  may be   motivated  \cite{Tseytlin:1985kh} by analogy with the first-quantized point-particle representation 
of the standard quantum field  theory effective action. For example, the expression for the 
torus part of the string partition function 
should effectively reduce, in the  point-like  limit, 
to the familiar relation between  the 1-loop  quantum effective action and minus the
 particle path integral on a  circle, 
$\Gamma_1 = \ha \log \det ( - \del^2 + ...) = - Z_{\rm part.}$, with $Z_{\rm part.} = \int_0^{\infty}{dt\ov 2t} \int [d x(\tau)]\, e^{- \int^t_0  d\tau(\frac{1}{2}\dot x^2 + ...)} $.\footnote{The integral over the Schwinger parameter $t$ can be understood as arising from the path  integral over the einbein upon gauge fixing the worldline diffeomorphism invariance (see for instance \cite{Polyakov:1987ez}).}

  \iffa 
F_mn  background (a  conf point)  and then Z_str= 1/Mobius  \int  d^D sqrt{ \det ( 1 +F) } 
Mobius SL2  volume  is naturally negative  (subtracting power divergence) ,   -2pi x 2pi = ads2 x U(1) 
 so that eff action 
 S_str =  \int  d^D sqrt{ \det ( 1 +F) }  is indeed -  Z_str.
Same  argument in closed  case is  unclear, but we may  indeed use the above two 
motivations   to suggest that *on-shell*  value of string eff action 
somehow is - Z_str (on-shell)    here as well  [here Mobius  is log divergent so 
precise argument is not really clear].
Anyway, I agree we may try to subscribe to this intuitive argument  adding above as some kind  of motivation and then  Z_inst sign is in line with F.
Trouble  is that above was for string but we need M2 brane   -- but as the two should be relate in a limit that suggests  same connection.
I will try to add something along these lines to sect 3
 \fi 

Before proceeding, let us comment on the interpretation of \rf{z}  expanded at large $k$   from the point of view of type IIA  string theory. 
Let us recall the relations  for the type IIA  string theory  coupling and tension
\ci{Aharony:2008ug}
 \ba
\la{2}
&\gs = \sqrt\pi\, \big({2 \ov k} \big)^{5/4}  {N}^{1/4}  =  \frac{\sqrt\pi\, (2\l)^{5/4}}{N} \ , \ \ \ \ \  \ \ \  \ \ \ \  \l= { N \ov k } 
\ , \\
&T = {1\ov 8\pi}  \frac{R_{10}^2}{\alpha'} = \gs^{2/3} \frac{\RL^{2}}{8\pi \alpha'}
= \frac{\sqrt{\l}}{\sqrt 2}\ , \ \ \qquad \qquad 
\ \     \frac{\gs^2 }{8\pi\, T }  ={\l^2 \ov N^2} =  {1\ov k^{2}} \ ,  \la{3}
\ea
where $\RL/2$  and $R_{10}/2$ are the  AdS curvature  radii in the 11d and 10d metrics.
Just like the $1\ov 2\sin {2\pi\ov k}$ prefactor in the Wilson loop  expectation value \rf{zz} 
sums up all the  leading   large $T$ terms  in the  higher genus  string corrections to the 
prefactor $ {1\ov \sqrt{2 \pi} }{\sqrt T \ov \gs}={k\ov 4 \pi}$ from the disk diagram \ci{Giombi:2020mhz},
similarly the prefactor in  $F^{\rm inst}_1$  in \rf{z}    may be written as 
\be 
  \frac{1}{\sin^{2} (\frac{2\pi}{k})} 
 =   \frac{1}{\sin^{2} (\sqrt{\pi \ov2}\frac{\gs}{\sqrt T})}  
 = {2 \ov \pi} {T\ov \gs^2}   +   {1\ov 3}  + {\pi  \ov 30 } {\gs^2\ov T}  
 + {\pi^2  \ov 378 } {\gs^4\ov T^2}  + ...\ , 
 \la{w}
 \ee
 where  the leading term is the contribution of the string tree level  2-sphere   diagram,  
 and corrections are  coming from higher genus  contributions. 
 To leading order  in  large tension $T$ at each order in $\gs^2$  we expect  that the exponential   with the  classical 
 instanton  action in \rf{z} will remain the same (the area of handles attached to 2-sphere  will be effectively negligible) 
 but the prefactor   will get corrected  order by order in $\gs$ as in \rf{w}. 
 This  picture is  corroborated  by  the  M2-brane   derivation  of \rf{z},\rf{w} below.

 \iffa 
 {\small
 \begin{verbatim}
 about higher genus it was menat to be indeed g_s  corrections to prefactors 
-- i.e. we compute determinants on a  surface with metric  depending on moduli 
and then integrate over moduli 
[to justify this modular integral  in static gauge we need again to start in conf 
gauge and then expect that main part of ghost det  will cancel against   longitudinal operator...
 but then why induced metric will depend on moduli ? ]
As for minimal surface it should be indeed same  but i am not totally sure 
how to justify this properly.--
same question -- we need induced metric to depend on moduli in dets  but 
why not in the action ? 
May be better  argument is that  if we  include that modular dependence in 
the action  exp [ - T (  a1 + a2(t_i)  ) ]   that may turn out to give only subleading 
corrections at leading  orders in 1/T ? --
 after all we qare after only  (g^2_s/T)^n  leading terms  and to capture 
*these* terms treating minimal surface as the same should be ok -- may be just add a 
comment like that ...
 \end{verbatim}
 }
 \fi

 The  leading   2-sphere term in \rf{w}   was recently discussed  in 
 \ci{Gautason:2023igo},
 where  its overall  factor was fixed  using relative normalization to other available data. 
 Like  in the WL  computation in \ci{Giombi:2023vzu}, here we will 
  not have  to rely on   indirect arguments to    fix  the tension dependence   of the prefactor 
  (all  UV   divergences  will cancel automatically) 
  but  will still need to address  the issue  of the 
   0-modes (appearing only from the string-level  fluctuations) 
    arguing for  the related overall factor of 2 following   \ci{Gautason:2023igo}.


\section{M2 brane  wrapped on $S^3/\ZZ_k$} 

\subsection{Classical M2 brane action in \adsz  background}

Let us  start with a review  of  some basic relations for the \adsz  background and 
the   classical  M2  brane action (see, e.g.,  \ci{Sakaguchi:2010dg,Giombi:2023vzu}). 

The 11d  metric is  ($n,m=1,2,3$;  \  $y\equiv y + 2 \pi$)
\ba
\la{3.1}
ds^{2}&=\frac{R^{2}}{4}ds^{2}_{AdS_{4}}+R^{2}ds^{2}_{S^{7}/\ZZ_{k}}\ , \ \ \ \ \ \ \ \ \ \ \ 
ds^{2}_{S^{7}/\ZZ_{k}} = ds^{2}_{\CP^{3}}+\frac{1}{k^{2}}(dy+kA)^{2},\\
ds_{\CP^{3}}^{2} &= \frac{(1+|w|^{2})\,dw^{n}\,d\bw^{n}-w^{n}\bw^{m}dw^{m}d\bw^{n}}{(1+|w|^{2})^{2}}, 
\la{3.3}  \\
\la{3.4}
A &= \frac{i}{2}(\bar\partial -\partial)\log(1+|w|^{2}) = \frac{i}{2}\frac{1}{1+|w|^{2}}(w^{n}d\bw^{n}-\bw^{n}dw^{n}).
\ea
We shall assume that AdS$_4$ has  Euclidean signature 
with  boundary $S^3$. 
This 11d   background is then 
supported  by $F_4 = dC_3 = -i \frac{3}{8}{\RL^3}\rm vol(AdS_4)$. 

\iffa 
The radius $R$ in units of the 11d  Planck  length $\lpl$ is related to the parameters $N$ and $k$ of the dual ABJM gauge theory 
by\foot{We are ignoring  higher order  corrections 
   to this relation (due to $N\to N - {1\ov 24} (k - { k^{-1}}) $  \ci{Bergman:2009zh})
as they will not be relevant  for  the leading   large $N$    M2 brane contribution considered below.} 
\begin{equation}
\left(\frac{R}{\ell_{\rm pl}}\right)^6 = 2^5 \pi^2 N k\,.
\label{R-map}
\end{equation}
\fi 

The  action for a (Euclidean)   M2 brane   in this    background  is given by \ci{Bergshoeff:1987cm,Bergshoeff:1987qx,deWit:1998yu,Pasti:1998tc,Claus:1998fh}
\begin{equation}
\IS = T_2\Big( \int d^3\xi \,  \sqrt{{\rm det}\, G} + i \int  C_3 \   + \ {\rm fermionic \ terms}\Big)\ , \qquad \ \  \ T_2 = \frac{1}{(2\pi)^2} \frac{1}{\ell_{\rm P}^3}\,.
\label{m}
\end{equation}
Here we  are interested in the   M2  brane configuration 
with $S^3/\ZZ_k$ world-volume, 
  that is wrapped on the 11d circle  $y$ of radius $R/k$ 
and on $\CP^1 \subset  \CP^3$.  This is the M2   uplift  of the IIA   string $\CP^1$ instanton of \ci{Cagnazzo:2009zh}.
The 
$\CP^{1}$  will be  chosen  as the  $w^{2}=w^{3}=0$
surface  in $\CP^3$.\footnote{This  corresponds to a particular 
 $\ha$-BPS  M2 brane  solution  
 discussed in \cite{Park:2020hgt}.}
We fix the world-volume reparametrization invariance   using  the static gauge:
we identify $(w_{1}, \bar w_{1}, y)$ with the 3 real 
world-volume coordinates $\xi^{i}=(u,v,s)$ according to 
\be
w^{1} \equiv z=  u+iv\ , \qquad \bw^{1} =\bar z= u-iv\ , \qquad y = s, \qquad  \qquad s\in(0,2\pi]  \ . \la{45}
\ee
As the $C_3$ potential has  only  the AdS$_4$   components 
the bosonic part of the corresponding  Euclidean M2  brane  action is  given by 
\be
\IS_{\rm cl} = T_{2}\, \int d^{3}\xi\, \sqrt{g}\ ,\la{46}
\ee
where 
$g_{ij}$ is the induced world-volume metric 
\be\la{47} 
ds^2_3=  g_{ij}d\xi^{i}d\xi^{j} = R^{2}\,\frac{dz\,d\bar z}{(1+|z|^{2})^{2}}+\frac{R^{2}}{k^{2}}\big[ds+k A(z,\bar z)\big]^{2}.
\ee
This is the metric of $S^{3}/\ZZ_{k}$  (for $k=1$ this is  the standard Hopf metric of $S^{3}$ with radius $R$).
The explicit form of the  metric   and the 1-form  $A= A_u du + A_v dv $  in the real basis $(u,v,s)$ is 
\ba
\la{3.8}
g_{ij} = R^{2}\,\begin{pmatrix}
 \kappa^{2}(1+v^2) & -\kappa^{2} u v & \frac{1}{k}A_{u} \\
 -\kappa^{2} u v & \kappa^{2}(1+u^2) &\frac{1}{k} A_{v}\\
 \frac{1}{k}A_{u} & \frac{1}{k}A_{v} & \frac{1}{k^2} \\
\end{pmatrix}, \quad A = \kappa(-vdu+udv), \quad \kappa \equiv (1+u^{2}+v^{2})^{-1} \ , \\
\la{3.9}
g^{ij} =\frac{1}{\kappa^{2}R^{2}}\, \begin{pmatrix}
1 & 0 & -k A_{u} \\
 0 & 1 & -k A_{v} \\
 -k A_{u} & -k A_{v} & k^2 \kappa \\
\end{pmatrix}, \qquad \sqrt g = \frac{R^{3}}{k}\kappa^{2}\ .\qquad  \qquad \qquad \ \ \ 
\ea
The resulting classical value of the action \rf{46}   is 
\be
\IS_{\rm cl} =  T_{2}R^{3}\, \text{vol}(S^{3}/\ZZ_k) = 
{1\ov k} T_{2}R^{3}\, \text{vol}(S^{3}) = {2\pi^2\ov k} \T_{2} \ . \la{410}
\ee
Here  the effective dimensionless tension is 
\be\la{412}
\T_{2} \equiv \RL^{3}  T_2={1\ov (2\pi)^{2}}   \frac{\RL^{3}}{\ell_{P}^{3}}
= {1\ov \pi} \sqrt{ 2 N k } \ , 
\ee
so that 
\be
\la{3.12}
\IS_{\rm cl} = 2\pi\, \sqrt\frac{2N}{k},
\ee
This is also the same as  the value of the classical  action of the string   world sheet  
wrapped on $\CP^1 $ in AdS$_4 \times \CP^3$ \cite{Cagnazzo:2009zh}, 
i.e.  $\IS_{\rm cl} = 2\pi\, \sqrt{2\l}$  (cf. \rf{3}). 

\subsection{Quadratic fluctuation Lagrangian}


Our aim will be to compute  the 1-loop  prefactor $\Z_1$ in  the   corresponding
 1-instanton contribution to the 
M2 brane partition function in  \rf{u}
\be 
Z^\inst_{\rm M2} = \Z_1 \,  e^{-  \IS_{\rm cl}} + ... \ . \la{413}  \ee
The factor $\Z_1$   will be expressed in terms of the  determinants of operators 
of the  bosonic and fermionic fluctuations  which will be functions of the 3d  coordinates 
$(u,v,s)$ in the static gauge   defined in \rf{45}. 

In this    static gauge we will have 8 real bosonic fluctuations:
4 in the AdS$_4$ directions and 4 in the 2 complex  transverse 
$\CP^3$ directions $w_2,w_3$. 
Fixing a $\kappa$-symmetry gauge
 (like in \ci{Bergshoeff:1987qx}), we will also have  8  fermionic 
fluctuations. 

Expanding the action \rf{m}, one   finds that the  fluctuations
 of $w_2$ and $w_3$   decouple 
and  their contributions  are the same.  
Considering, e.g., $w_2$    and setting (cf. \rf{3.8})\foot{Here $\kappa(u,v)$ is defined in \rf{3.8}.
We also rescale all fluctuation fields by  $\T_2^{-1/2}$.} 
\be
w_{2}(u,v,s) = \kappa^{-1/2}\, \phi(u,v,s) \ , 
\ee
we get   for the   corresponding quadratic fluctuation Lagrangian
\ba
\la{4.4}
\mc L_{2} (\phi)  &= \frac{R^{2}}{2}\sum_{i,j=1}^{3}g^{ij}D_{i}\bar\phi\,D_{j}\phi 
- \bar{\phi }\phi- \frac{i}{2} k ( \bar{\phi } \del_s \phi
- \del_s  \bar{\phi}\,\phi) \  , \ \\
& D_{i}\phi =( \partial_{i}-i \, A_{i})\,\phi, \qquad
D_{i}\bar\phi = (\partial_{i}+i \, A_{i})\,\bar\phi  \ . \la{455}
\ea
Here $\del_i= (\del_u, \del_v, \del_s) $ 
and  $A_{i}=(A_u,A_v,0)$ is the 3d  gauge potential  
with $A_u, A_v$ defined in  (\ref{3.8}).

As $s$  is a periodic coordinate   we may  interpret the corresponding  3d  action $\int d^3 \xi \sqrt g \ \mc L_{2}$ as a 2d action 
 for an infinite tower of the 
Fourier modes of $\phi$  by setting 
$\phi(u,v,s) = \sum_n \phi_{n}(u,v)\,e^{ins}$. 
This 2d action will be defined on $\CP^1$  with the metric $g_{ab}$ of a 2-sphere of radius $R/2$
 (the  first term in \rf{47}). The corresponding 
  Lagrangian for a tower of 2d charged  massive  complex 
   scalars $\phi_n$  on the 2-sphere  coupled to the background 2d 
   abelian  gauge  field potential 
   $A_a$  is then (here $a,b=1,2$  label  the $u,v$ directions)
\ba
\mc L_{2} (\phi_n) 
 = &\frac{R^{2}}{2} \Big[
\sum_{a,b=1}^{2}g^{ab} D_{a}\bar\phi_{n}\, D_{b}\phi_{n}+M^{2}\, \bar\phi_{n}\phi_{n}
\Big]\ ,  \la{4.5} \\
D_{a}\phi_n   =& (\partial_{a}-i\,q A_{a})\phi_n  \ , \qquad  D_{a}\bar 
\phi_n   = (\partial_{a}+ i\,q A_{a})\phi_n \ , \\
 q =& 1+nk ,  \qquad\ \ \ \ \ \ \ 
R^{2}M^{2} = -2+2nk+n^{2}k^{2} \ . \la{54} 
\ea
The fluctuations in the AdS$_{4}$  directions   may be 
represented  by 4 real 3d  massless scalars 
 $\eta^{r}$  ($r=1,2,3, 4$) with the 3d  Lagrangian
\be
\mc L_{2}(\eta)  =\frac{R^{2}}{2} \sum_{i,j=1}^{3}g^{ij} \del_{i} \eta^{r}\, \del_{j}\eta^{r}.
\ee
The corresponding 2d  Lagrangian for the tower of the Fourier modes of 
$\eta^r(u,v,s) = \sum_n \eta^r_{n}(u,v)\,e^{ins}$ 
(with $\eta^r_{- n}= \bar \eta^r_{n}$)  then  has a similar form to \rf{4.5}  
\ba
\mc L_{2}(\eta_n) 
 =& \frac{R^{2}}{2}\Big[\sum_{a,b=1}^{2}g^{ab} D_{a}\eta^{r}_{-n}\, D_{b}\eta^r_{n}+M^{2}\, \eta_{-n}^r\eta^r_{n}\Big], \la{421} \\
\la{4.7}
D_{a}\eta^r_{n} =& (\partial_{a}-i q A_{a})\, \eta^r_{n}, \qquad q= nk , \qquad R^{2}M^{2} = n^2k^2  \ .
\ea
Let us   comment on  the explicit 
 form of the background    metric and  gauge field  in the 2d actions corresponding to \rf{4.5},\rf{421}. 
The metric $g_{ab}$  obtained by the   restriction of the 
 induced   metric  to the $u,v$ subspace is  the metric of  the 2-sphere 
 with radius $L=\frac{R}{2}$:
\be\la{yz}
g_{ab}d\xi^{a}d\xi^{b} = R^{2}\frac{du^{2}+dv^{2}}{(1+u^{2}+v^{2})^{2}}
= L^{2}(d\theta^{2}+\sin^{2}\theta\, d\varphi^{2}), \qquad u+iv = \tan\frac{\theta}{2}\, e^{i\varphi} \ , \qquad L=\frac{R}{2} \ .
\ee
The  gauge potential 
 $A$ in (\ref{3.8})   written in these angular   coordinates $\theta, \varphi$ reads 
\be
A = \frac{1}{2}(1-\cos\theta)\, d\varphi,\ \ \ \ 
F= dA =  \frac{1}{2}\sin\theta\, d\theta\wedge d\varphi \ , \ \ \ \
\frac{1}{2\pi}\int_{S^{2}} F = 1 \la{425} \ . 
\ee
It  may be interpreted as  a field of a   unit-charge 3d monopole 
placed at the center of  a unit-radius $S^{2}$. 

The  fluctuation operators  in \rf{4.5} \rf{421}  are thus the   standard 
2nd order operators
on $S^{2}$ of  radius $L$ in the   magnetic monopole    background 
\ba
\la{4.12}
\Delta= -D^{2}+M^{2},  \qquad\qquad D_{a} = \partial_{a}-i q A_{a}\ . 
\ea
Measuring the masses  in terms of the radius $L=R/2$  of $S^2$  
 we  thus get the  following  bosonic spectrum: 2 towers of complex $\phi_n$ modes  and 
4 towers of  $\eta_n= \bar\eta_{-n}$ modes  with 
\be \la{tt}
\phi_n: \ \ \    m^2 \equiv L^2 M^2 = -{3\ov 4} + {1\ov 4} ( 1  + nk)^2 \ , \ \ \ \ 
q= 1 + nk ; \qquad \ \ \ 
\eta_n: \ \ \ \ m^2  = {1\ov 4} ( nk)^2 \ , \ \ \ \ 
q= nk\ . 
\ee

The M2 brane   action in \adsz   background is related to the type IIA  string in the corresponding AdS$_4 \times \CP^3$   background   by the double   dimensional reduction  \ci{Duff:1987bx}. 
Indeed,  the   $n=0$  parts    of the 2d fluctuation  Lagrangians 
\rf{4.5},\rf{421}   are equivalent to the ones in the type IIA string case  in  \ci{Gautason:2023igo}. 

This    relation is even more direct in the fermionic sector 
as the fermionic fields  in the M2  brane  and the type IIA GS  string actions
are in direct correspondence (their  components  are essentially 
the same, the only difference is  due to the M2 brane  fields depending on the extra coordinate $s$). 

It is thus   straightforward to reconstruct the quadratic part of the 
 M2  brane  fermionic  action  from its  lowest KK  level $n=0$ term, i.e.
 by starting   with the fermionic  part of the type IIA superstring action 
 used in \cite{Gautason:2023igo}.
 The detailed structure of the quadratic fermionic  Lagrangian  in the string 
  case\footnote{We thank the authors of \cite{Gautason:2023igo} for kindly sharing their unpublished notes on the derivation of the fermionic fluctuation part of the type IIA superstring  action in the background of the $\CP^1$ instanton.} 
  shows 
  that it   is equivalent   to the sum of   2d fermionic  terms $\bar \psi \mc D \psi$
   where $\mc D $ is the standard   2d 
 Dirac operator on the 2-sphere  of radius $L= R/2$ 
 \rf{yz}  in  the monopole background \rf{425} 
 with a particular   mass term
 (\cf (\ref{4.12}))
\ba
\la{4.13}
\mc D &= i\,\slashed{D}+M_{1}\sigma_{3}+M_{2},\qquad  \qquad \slashed{D} = \sigma^{\rm a }e{_{\rm a}^{a}}(\partial_{a}
+\frac{i}{2}\omega_{a}\sigma_{3}-i q  A_{a}) \ . 
\ea
Here 
 $e{_{\rm a }^{a}}$ is the inverse zweibein on the sphere (${\rm a} = 1,2$), 
  $\omega_{a}$ is the 2d spin connection,  and $\sigma_i= (\sigma_{\rm a}, \sigma_3)$  are the Pauli matrices. 
  The explicit  values of the  dimensionless  mass parameters are 
\be
\la{4.14}
m_{1} \equiv L M_1=   -\frac{1}{4}(u-u'),\ \  \qquad m_{2} \equiv L M_2=  -\frac{1}{4}-\frac{3}{4}uu',\qquad u,u'\in\{1,-1\},
\ee
where $u,u'$ represent 4  independent sign  factors  
arising from 10-d Gamma matrices in a  suitable representation. Thus  
 one finds  8 fermionic modes organized as 2d fermionic  fields  with 4 choices of mass  parameters in \rf{4.13} 
$m_1= (-\ha, \ha, 0,0), \ m_2= (\ha, \ha , -1, -1)$. 
In addition, the values 
of the charges  are $q=(1,-1,0, 0)$  \cite{Gautason:2023igo}.


To generalize this to the M2  brane  case, again expanding the fermions in the Fourier  modes
$\theta (u,v,s) = \sum_n \theta_n (u,v) \, e^{i n s}$  in the 3rd  direction $s$ (with $s$   identified with the 
 11-th  direction $y$ in the static  gauge \rf{45} we use), i.e.  
we need to account for the contribution  of the  corresponding 
term 
\be  \la{p}
\bar \theta (\Gamma^{A }  E^y_{A }\partial_{s}+ ...) \theta=
\bar \theta (\Gamma^{11 }  E^y_{11 }\partial_{s}
+  \Gamma^{a }  E^y_{a }\partial_{s} +   ...) \theta \, ,  \ee
  in the supermembrane  action  \ci{Bergshoeff:1987cm,Bergshoeff:1987qx}.\foot{The quadratic fermionic term in the 
M2  brane   action is built  using  the   gravitino 
covariant derivative in 11d supergravity  and is straightforward  to analyze, 
given that here $y= x^{11}$ is an isometric direction.}
As  $\Gamma_{11}$  gets expressed  in terms of $\sigma_3 $ (times a unit matrix)   we   learn that the $M_1$ term in \rf{4.14} 
 gets a  shift  (due to $\del_s \theta \to    in \theta_n$)   while   $M_2$  stays the same, i.e.\foot{Here $n=0, \pm 1, \pm 2, ...$  so the sign of the shift is not important.} 
$
\Delta m_{1} =  -\frac{1}{2}{nk}, \  \Delta m_{2} =0 .
$
In addition, the presence of the off-diagonal $A dy $ term in the 11d metric 
\rf{3.1},   and thus  in the  corresponding 
vielbein,   implies  also 
that the $\del_s$ term in \rf{p}   leads to a shift  of the coefficient of 
the $A$-term in the covariant derivative in \rf{4.13}, i.e. to the  shift  of  the charge 
$\Delta q =  nk$.\foot{Let us note that the 
 reason why the 
 mass $m_{1}$   gets a  shift $\ha nk $
  while the charge gets a shift of   $nk$ (that seems to contradict the usual KK  intuition)  has to do with the fact 
  that as follows from the  structure 
  of the 11d metric the relevant 
  $S^3/\ZZ_k$    
     subspace has the radius of
  the 2-sphere being   $L=R/2$  and that translates   into the  
  factor 2 in the relative $nk$  shift of charge  compared to mass. 
  The same is seen  also in the bosonic spectrum in \rf{tt}.}


Thus  we find that   the Lagrangian for the  tower of the 2d fermionic modes  originating from the quadratic fermionic part of the 
M2   brane action can be represented  by 
a collection of 4   2d fermionic fields   with the Dirac-like operators \rf{4.13}  where 
the parameters depend on $k$ as  
\be
\la{444}
m_{1} =   -\frac{1}{4}(u-u') -\frac{1}{2}{nk},\ \  \qquad m_{2} =  -\frac{1}{4}-\frac{3}{4}uu',\qquad \ \ \  q= - 2 m_1 \ . 
\ee

\section{Determinants  of   operators on $S^{2}$ in  monopole background}


\subsection{Spectra  of  operators and formal spectral sums}

For a  massless  scalar field of charge $q$  on $S^{2}$  
in the field of a monopole  normalized as in   \rf{425}
the spectrum of the  corresponding 
 Laplace operator \rf{4.12}  was  found in   \cite{Wu:1976ge}. Its eigenvalues (normalized  to  the radius $L$  of the sphere, i.e.  multiplied by $L^2$)  and  
 degeneracies  are given by 
\ba
\lambda_{\ell} &= \ell(\ell+1)-\frac{q^{2}}{4},\qquad   \qquad \ell-\frac{|q|}{2}=0,1,2,\dots, \qquad \text{deg}\lambda_{\ell} = 2\ell+1 \ . 
\ea
Inclusion of  mass term  in the operator   can be done  by 
 the obvious   shift $ \lambda_{\ell} \to \lambda_{\ell}  + m^2, \ \   m\equiv L M$.  Then the 
formal (unregularized)   expression for the corresponding  determinant 
 may be written as 
\be
\la{5.2}
\log\det \big[L^{2}(-D^{2}+M^2)\big] = \sum ^\infty_{\ell=\frac{|q|}{2}}(2\ell+1) \log\Big[\ell(\ell+1)-\frac{q^{2}}{4}+m^{2}\Big] = \sum^\infty_{\ell=\frac{|q|+1}{2}}2\ell\, \log\Big[\ell^{2}-\frac{1}{4}-\frac{q^{2}}{4}+m^{2}\Big].
\ee
 To obtain the last relation we redefined the summation index  $\ell$ by 
  $\frac{1}{2}$; such a shift  
is allowed assuming  one adopts  a spectral regularization
  like spectral zeta function (as we will  do  below).
  Introducing  the   notation 
\be
\la{5.3}
s_{p}(\mu) \equiv \mathop{\sum^\infty_{\ell=p, \  \ell^{2}\neq \mu}}
2\ell \log(\ell^{2}-\mu)\ , 
\ee
where $p$  and $\mu$ are some parameters  and  $\ell - p$ takes  non-negative  integer  values, eq.  \rf{5.2}   may be written as 
\be
\la{5.4}
\log\det\big[L^{2}(-D^{2}+M^{2})
\big] = s_{p}(\mu)\ ,
\qquad  \qquad p = \frac{|q|+1}{2}\,,\ \  \  \qquad \mu = \frac{1}{4}+\frac{q^{2}}{4}-m^{2}\ .
\ee

Turning to the 2d  spin $\ha$  charged   field case, the 
corresponding  massless  Dirac operator $i\slashed{D}$  defined in \rf{4.13}
 has  eigenspinors  with the following  eigenvalues (normalized again to  the radius $L$ of the sphere)  and degeneracies  
\cite{Dolan:2003bj,Dolan:2020sjq} \be\la{556}
\lambda_{\ell} = \pm\sqrt{\ell^{2}-\frac{q^{2}}{4}},\ \ \ \ 
 \qquad \ell-\frac{|q|}{2} = 0, 1, 2, \dots, \qquad\ \  \text{deg}\lambda_{\ell}=2\ell.
\ee
For the minimal value $\ell=\frac{|q|}{2}$ (assuming $|q|\ge 1$), we
 get  $|q|$ zero modes with definite chirality 
 (or  the eigenvalue of  the $\sigma_{3}$  matrix)
equal to the sign of $q$, consistently with the Atiyah-Singer index theorem on the 2-sphere \cite{Deguchi:2005qd}.

In the  case of the massive  operator  $ i\slashed{D} + M_1 \sigma_3 + M_2$ 
 in \rf{4.13} 
 this  spectrum leads  to the following expression  for the  determinant 
  (here 
  $m_a = L M_a$)
\be
\la{5.6}
 \log\det\big[L(i\slashed{D}+M_{1}\sigma_{3}+M_{2})\big] =|q|\,\log\big|\sign(q)\, m_{1}+m_{2}\big|
+\sum_{\ell=\frac{|q|}{2}+1}^{\infty}2\ell\log\Big(\ell^{2}-\frac{q^{2}}{4}+ m^2_1 - m^2_2 \Big) .\ \ \ \ 
\ee
Here the  first term
 represents the contribution of the 0-mode of $\slashed{D}$  which has a definite chirality equal to  $\sign(q)$, as
  explained above.\footnote{ 
More details on the explicit eigenspinors of the Dirac operator on $S^{2}$ are discussed in \cite{Abrikosov:2002jr}. The fermionic 
   spectral problem  on $S^{2}$ in the monopole background is treated also  in \cite{Borokhov:2002ib,Pufu:2013vpa,Dyer:2013fja}.
}
That the  eigenvalues of all other modes  contain the effective mass-squared parameter $m^2_1 - m^2_2$  (cf. \rf{5.2})
follows from the direct evaluation of the   determinant in \rf{5.6}  or  can be seen  from  ``squaring'' the   
first-order operator as in \rf{5.17}.

The fermionic counterpart $ \wt s_{p}(\mu; {\sfm })$ of the function $s_{p}(\mu)$ in (\ref{5.3})  may be defined as 
\be 
\wt s_{p}(\mu; {\sfm }) \equiv   2 p\,\log{\sfm }+s_{p+1}(\mu)\ , 
 \la{op}\ee
so that (cf. \rf{5.3},\rf{5.4})
\ba
\la{5.7}
& \log\det\big[L (i\slashed{D}+M_{1}\sigma_{3}+M_{2})\big] = \wt s_{p}(\mu; {\sfm }) \ , \\
\la{577}
&  p = \frac{|q|}{2}, \qquad \mu = \frac{q^{2}}{4}-m_1^2 + m^2_2 ,    \qquad {\sfm } = \big|\sign(q)\, m_{1}+m_{2}\big|.
\ea


\subsection{Computing determinants  using spectral $\z$-function}
The  determinant of an elliptic 2nd order operator $\Delta$ can be expressed in terms of the  spectral $\z$-function   $\z_\Delta (z) =\sum_\ell  \lambda^{-z}_\ell$, \ $\lambda_\ell\neq 0,$   as 
\be \la{67}
\log\det\Delta = - \z_\Delta (0)\, \log ( \Lambda^2  L^2) + 
 (\log\det \Delta)_{\rm fin}\ , \qquad \ \ \ 
 (\log\det \Delta)_{\rm fin}  = -\zeta'_\Delta(0)
\ ,  \ee
where $\Lambda$ is a  2d   UV cutoff. 
In particular, for the bosonic operator in \rf{5.4}    we get 
\ba\la{513}
\zeta_\Delta(\s) &= \sum_{\ell=p}^{\infty}2\ell\,(\ell^{2}-\mu)^{-\s}\ .
\ea
This  can be computed by expanding in series of $\mu$
\ba
\la{5.13}
\zeta_\Delta(\s) &= \sum_{\ell=p}^{\infty}2\ell^{-2\s+1}\,(1-\mu\ell^{-2})^{-\s} = 2\,\sum_{k=0}^{\infty}\binom{-\s}{k}(-\mu)^{k}\sum_{\ell=p}^{\infty}\ell^{-2\s-2k+1}\lp
= 2\,\sum_{k=0}^{\infty}\binom{-\s}{k}(-\mu)^{k}\,\zeta(2\s+2k-1,p) \ , 
\ea
where $\zeta(x,a)$   is the Hurwitz $\zeta$-function.
As a result, 
\be
\la{5.14}
\zeta_\Delta(0) = -\frac{1}{6}+\mu +p(1-p) \ . 
\ee
In particular, for  the values  of $p$  and $\mu$ 
corresponding to the bosonic operator in  (\ref{5.4})  we get  
\be
\la{5.15}
\zeta_\Delta(0) = \frac{1}{3}- m^2,  \qquad \qquad   m^2=L^2 M^2\ . 
\ee
This   matches    the value   of the second 
Seeley coefficient $B_2$  that controls the  log UV divergent part of 
$\log \det $ of the   operator $- D^2 + X $  on  $S^2$ in the heat kernel regularization:\foot{Note that $B_2$
 does not  depend on  2d gauge field $A$   and indeed the dependence on $q$  cancels in \rf{5.14}.}
$B_2= {1\ov 4\pi} \int d^2 \xi \sqrt g \ (\frac{1}{6}{\cal R} - X) = \frac{1}{3}-L^{2}M^{2} $  where  $X= M^2$ and 
${\cal R} = {2\ov L^2}$   is the curvature of $S^2$
with  area $4\pi L^2$. 

In the  case of the fermionic operator in \rf{5.6},\rf{5.7} 
we need to  account for the fact that for the lowest value of $\ell=p$  there is just one  and not two eigenvalues  with multiplicity $p$
 (cf. \rf{op}).
Then instead of  (\ref{5.14}) we get for  the $\zeta$-function corresponding to the operator in 
 \rf{5.6}\foot{Equivalently,  we are to add $p$ to the case 
 with $p'=p+1$, i.e. $p'(1-p')  + p= p(1-p) -   p = - p^2$.}
\be
\la{5.16}
\zeta_\Delta(0) =-\frac{1}{6}+\mu + p(1-p) -  p
 = -\frac{1}{6}- m^{2}, \ \ \ \qquad  m^2= L^2 (M_1^2 - M_2^2 )\ .
\ee
This  again  is in correspondence  with 
the   value of the $B_{2}$ coefficient. 
 To see this     note that $\log \det \mc D = \ha \det \Delta$  where 
 $\Delta $  is the corresponding ``squared'' Dirac operator (here
 $\tilde {\mc D}=S\mc D S^{-1} $ where 
  $S$ is an appropriate spinor rotation matrix;  we suppress spinor indices)
 \be
\la{5.17}
\hat \Delta  =\tilde {\mc D}\mc D =  -D^{2}+\frac{1}{4}{\cal R}  +  \ha   i\,q\,\epsilon^{ab} F_{ab}\sigma_{3} +M_1^{2} - M_2^2
\ . \ee
As  this operator  is of the general form $-D^2 + X$ 
 we  have $B_2= {1\ov 4\pi} \int d^2 \xi \sqrt g \ \tr (\frac{1}{6}{\cal R} - X) $  and thus   $B_2 = 2\,  {1\ov 4\pi} \int d^2 \xi \sqrt g  \, [ 
\frac{1}{6}{\cal R} -(\frac{1}{4}{\cal R}+M_1^2 -M_2^2)] = 
2 [-\frac{1}{6}-L^{2}(M_1^2-M_2^2)]$  which is consistent with \rf{5.16} 
 corresponding to the 1st order operator in \rf{5.6}. 
    
    Let us note that in general  $\zeta_\Delta (0)$ represents the  regularized  total number
     of all non-zero modes.  In the above discussion we were assuming that 
    the values of parameters are such that  there are no  zero modes.
    In the  case when there are $N_0$  zero modes 
    one should   find that   $B_2=  \zeta_\Delta(0) + N_0$.
       For a particular system of operators   discussed  in the next section we will have equal  numbers of 
       the bosonic and fermionic  zero modes  and  thus 
       the sum of $\sum_i (-1)^{F_i} B_2(\Delta_i) $
         and $\sum_i (-1)^{F_i} \zeta_{\Delta_i}(0)$  will still be equal.

Let us now  compute 
 the finite   part  $(\log\det \Delta)_{\rm fin}$  in \rf{67}.
In the bosonic  case  we find from (\ref{5.13})  (here $\psi(x) = (\log \Gamma(x))'$)
\be
\la{5.19}
\frac{d}{d\mu}  (\log\det \Delta)_{\rm fin}
 = \sum_{k=0}^{\infty}\frac{2}{(2k)!}\psi^{(2k)}(p)\, \mu^{k} = \psi(p+\sqrt\mu)+\psi(p-\sqrt \mu).
\ee
The integration constant is fixed by the value of $\zeta'_\Delta(0)$ at $\mu=0$ which  is also obtained from (\ref{5.13}) 
\be
\zeta_\Delta(z)\big |_{\mu=0} = 2\zeta(2z-1,p) \qquad\to\qquad \zeta_\Delta'(0)\big|_{\mu=0} = 4\zeta'(-1,p).
\ee
Then from  (\ref{5.19}) we find (see, e.g., \ci{Allen:1983dg} 
and  \cite{Gautason:2023igo})
\be
\la{5.21}
  (\log\det \Delta)_{\rm fin}
   = -4\,\zeta'(-1,p)+\int_{0}^{\mu}dx\ \big[\psi(p+\sqrt x)+\psi(p-\sqrt x)\big]
   \equiv s_p(\mu),
\ee
where in  what follows we will define $s_p(\mu)$ in \rf{5.3},\rf{5.4} by  this 
regularized expression. 

Similarly, 
in the case of the  fermionic operator in \rf{5.6}
we obtain the finite part of  $(\log\det[L (i\slashed{D}+M_{1}\sigma_{3}+M_{2})])_{\rm fin}=\wt s_{p}(\mu; {\sfm }) $
by using \rf{5.7} and \rf{op}  where now  $\wt s_{p}(\mu; {\sfm })$ is  expressed  in terms  of 
 $s_{p+1}(\mu)$ 
 in its   regularized form \rf{5.21}, i.e.  
\be \wt s_{p}(\mu; {\sfm }) = s_{p+1}(\mu) +  2p\,\log{\sfm }  =
 s_{p}(\mu)  -2p \log(p^{2}-\mu) + 2p\,\log{\sfm } \ . \la{666}
\ee
Here we used that according to the definition in \rf{5.3}  
one  finds   that   $s_{p}(\mu) = 2p \log(p^{2}-\mu)  + s_{p+1}(\mu) $
(the same  relation follows also  from   (\ref{5.21})). 
Thus the bosonic and fermionic  log det contributions 
with the same $p$ and $\mu$   have  the same non-trivial  part \rf{5.21}, i.e.  
 differ  only by the  two logarithmic terms  in \rf{666}.

\section{One-loop  instanton prefactor in the   M2 brane partition function}

\subsection{Summary of the spectral data}

We are now ready   to combine  the   above results  to compute the prefactor $\rm Z_1$ in \rf{4.13}.
Let us start with   summarizing in Tables 1  and 2 
the data about the   bosonic and fermionic
fluctuation  spectra 
 in \rf{tt}  and \rf{444}  and the corresponding parameters $p$  and $\mu$
 in the operators  in \rf{5.4}  and \rf{577}. 
 The type IIA  string ($n=0$)  part of the spectral   
  data  in Tables \ref{tab:bos} and \ref{tab:fer} was found earlier in 
 \cite{Gautason:2023igo}.
\begin{table}[H]
\be
\def\arraystretch{1.3}
\begin{array}{cccccc}
\toprule
& \text{\# real scalars} & m^{2} & q & \mu & p  \\
\midrule
{\rm AdS}_{4} & 4 & \frac{1}{4}n^{2}k^{2} &  nk & \frac{1}{4} & \frac{1+|n|k}{2} \\
\midrule
\CP^{3} & 2+2 & \frac{1}{4}(-2+2nk+n^{2}k^{2}) & \pm(1+nk)  & 1 & \frac{1+|1+nk|}{2}
 \\
\bottomrule
\end{array}\notag
\ee
\caption{Bosonic spectrum. The parameters $\mu$ and $p$ are defined in eq.~(\ref{5.4}).\la{tab:bos}}
\end{table}
\noindent

\begin{table}[H]
\be
\def\arraystretch{1.3}
\begin{array}{cc|cc|cc|ccccc}
\toprule
u & u' & m_{1} & m_{2} & m^{2} = m_{1}^{2}-m_{2}^{2} & q & p & \mu  &  {\sfm } _{n>0} & {\sfm }_{n<0} & {\sfm }_{n=0} \\
\midrule
+1 & -1 & -\frac{nk+1}{2} & \frac{1}{2} & \frac{nk(nk+2)}{4} & {1+nk}   & \frac{|nk+1|}{2}& \frac{1}{4} &  \frac{nk}{2}     &   -\frac{nk}{2}-1 & 0  \\
-1 & +1 & -\frac{nk-1}{2} & \frac{1}{2}  & \frac{nk(nk-2)}{4}& {-1+nk}    & \frac{|nk-1|}{2}& \frac{1}{4} & \frac{nk}{2}-1    & -\frac{nk}{2} & 0 \\
+1 & +1 & -\frac{nk}{2} & -1                    & \frac{k^{2}n^{2}}{4}-1      & {nk}  & \frac{|n|k}{2} & 1 & \frac{nk}{2}+1      & -\frac{nk}{2}+1 & 1 \\
- 1 & - 1 & -\frac{nk}{2} & -1                    & \frac{k^{2}n^{2}}{4}-1      & {nk}  & \frac{|n|k}{2} & 1 & \frac{nk}{2}+1      & -\frac{nk}{2}+1 & 1 \\
\bottomrule
\end{array}\notag
\ee
\caption{Fermionic spectrum. The parameters $\mu$, $p$ and $\sfm$ are defined in eq.~(\ref{577}). \la{tab:fer}}
\end{table}
\noindent
%
The 4 rows  in Table 2  correspond to the parameters  of the 
complex 2-component 2d  fermions (one  complex 2d fermion represents 
the same  number of degrees of freedom as  one complex scalar). 
 The  values of the  parameter $\mu$  in \rf{577} and the charge $q$ 
in the fermionic  spectrum  
are the same as in the bosonic spectrum (were $\mu$ is defined in \rf{5.4}).
This  should  be an  indication of an underlying 2d supersymmetry which  should be 
related to the fact that the M2 brane instanton 
background preserves half of the  target space supersymmetry.

Before proceeding, let us comment  on the zero modes  in the spectrum that are  not included in the determinants  computed via spectral $\zeta$-function 
and thus should be treated separately.


In the bosonic sector, zero modes may  appear when $\ell^2 =\mu$ (cf. \rf{5.3},\rf{5.4}). From the data in Table  1  where $\mu=1$ or $\mu= {1\ov4}$ 
and  given that $\ell$  starts with $p$  we conclude (assuming as always  that $k>2$) 
that this is possible only 
for $n=0$ modes with $\ell=\ha$ for  four AdS$_4$ fluctuations
 and $\ell=1$  for four  $\CP^3$ fluctuations. Taking into account the degeneracy $2\ell$ of each mode   
  we get in total $4 \times (1 +2)=12$  real bosonic zero modes.
  
   The same number of 12 real   zero modes is found also 
    in the fermionic sector: again zero modes appear  only for $n=0$    and thus 
     their count is equivalent to the one in 
     \cite{Cagnazzo:2009zh,Gautason:2023igo}.
    For the fermionic determinant in \rf{5.6},\rf{577}
    a  zero mode is found if  ${\sfm }=0$   or 
    $\ell^2 =\mu= {q^2\ov 4} - m^2$  for $\ell\geq p={|q| \ov 2}$. 
    We thus get   $2+2$ modes   from the first  two lines in Table 2
        and another $4+4$   from the last two  lines.

The four  bosonic  AdS$_4$ 0-modes correspond
 to the point-like  position of the instanton in AdS$_{4}$.
 As for the $\CP^3$  modes, their count  is related to   earlier  discussions 
 of the instanton 0-modes  in  the standard $\CP^{\rm N}$   sigma model on  2-sphere  
 \cite{Polyakov:1975yp,DAdda:1978vbw,Din:1979zv,Narain:1983in} 
 as follows. 
 
 Considering a $\CP^1$ instanton   in the standard $\CP^3$   sigma model 
 one finds 2 ``longitudinal'' fluctuations   and 4 ``transverse'' fluctuations. 
 If we were to quantize the AdS$_4\times \CP^3$ string in the conformal 
 instead of the static gauge
 then the  contribution of the longitudinal fluctuations    would cancel against   that   of the  conformal gauge ghost operator on 2-sphere. The latter  
  has 6 real   $SL(2,\CC)$ M\"obius zero modes (corresponding to the conformal Killing vectors on $S^2$).  The same six 
    0-modes are found also for the longitudinal   fluctuation operator  whose  contribution  cancels   against the  ghost determinant one. This cancellation extends also  to the  
    0-mode factors, leaving the static gauge result where
     only the   contributions of the transverse   fluctuations 
      are present.\foot{Note that in the static  gauge in string theory   there is no 
 factor  of the  M\"obius  volume as it cancels against the 0-mode factor in the  
 determinant of the longitudinal modes (cf. also a   discussion in \cite{Giombi:2020mhz}).}
  Thus  the total number 
  of the  0-modes  in the $\CP^3$    string sigma model  in the conformal gauge is $6+8=14$. 
  This agrees  with the  number $4\rm N +2 $ of 0-modes 
  in the $\CP^{\rm N}$  sigma model expanded near $\CP^1$ instanton
  \cite{Perelomov:1987va}: $ (4\rm N+2)\big|_{N=3} =14$.


  \iffa \foot{
In more detail,  the one-instanton solution 
 in $\CP^{N}$  is represented by a meromorphic function in suitable complex coordinates and has $2(N+1)$ real parameters once we fix the asymptotic value at infinity \cite{Perelomov:1987va}.
This is a generic point of $\CP^{N}$ and corresponds to further $2N$ real parameters, for a total of $4N+2$ parameters, \ie  14 for $N=3$.
Our static gauge choice fixes the asymptotics and leaves only the first set of $2(N+1)$ parameters, \ie 8 for $N=3$. 

An alternative gauge fixing would be conformal gauge including ghosts. 
Notice that this is an incomplete gauge choice leaving unfixed conformal symmetry, see App. A2 of \cite{Giombi:2020mhz}.
In this case, the 2 extra longitudinal fields in $\CP^{1}$ will have
non-constant modes canceling against corresponding non-constant modes of ghosts. Zero modes of these fields will be also in correspondence
and are 6 modes corresponding to $3\times 2$   M\"obius (conformal Killing) modes on $S^{2}$. 
The M\"obius group has volume $\text{vol}(SL_{2}(\mathbb C))$ which is log divergent. The result is usually
written with a factor $1/\text{vol}(SL_{2}(\mathbb C))\times  \det(\text{ghost})$ which is formally zero either because of the volume factor
or ghost operator zero modes.

As a final comment, let us remark that for the M2 brane conformal gauge  will not be completely conformal:   we need to assume that M2  is wrapped on $x_{11}$  
so  this  will be like partial static gauge in this regard;  but then we can  fix 2 remaining reparametrizations  by  conditions of  induced metric to be conformally flat.
\fi

\subsection{The total contribution of the fluctuation determinants}

Let us now   sum up   the log det 
contributions of all  2d fluctuation fields  with the spectral data  in Tables 1 and 2.
 The corresponding   1-loop   correction  is (cf. \rf{67})
\be\la{777}
 \Gamma = \Gamma_B - \Gamma_F=\ha\sum^\infty_{n=-\infty} \sum_i (-1)^{F_i} \log {\rm det}' \Delta_i
=  - \zeta_{\rm tot}(0) \log (\Lambda L) - \ha \zeta'_{\rm tot}(0)\ , \ee
where each term is  counted as coming from a real 2d field. 
We  are   not including here the  contribution of  the  bosonic  and fermionic modes 
that appear only for $n=0$; their contribution  produces just a constant    that can be found 
as   suggested in \ci{Gautason:2023igo} (see also below).

Let   us first  compute  the value of the 
 total coefficient $\zeta_{\rm tot}(0)
= \sum^\infty_{n=-\infty} \zeta_{{\rm tot},\,  n}(0)$ of the logarithmic  divergence in \rf{777}. 
Applying the expressions in \rf{5.15}  and \rf{5.16}  we  find that for a fixed 
level $n$   the result is $\zeta_{{\rm tot},\,  n}(0) = 2-2nk$.
Here the $n=0$ is the non-vanishing  type II 
GS string theory  coefficient  which is, in general,  given by the Euler 
number of the  2-surface  \ci{Drukker:2000ep,Forini:2015mca,Giombi:2020mhz}. 
This is  2 for $S^2$ in the present case   while it  was 1  in  the Wilson loop   case in 
 \ci{Giombi:2020mhz,Giombi:2023vzu}  corresponding to disk topology. 
 Summing over $n$  using  an  analytic regularization  (Riemann $\zeta$-function)  we then find that 
 \be \la{177}
 \zeta_{\rm tot}(0)
= \sum^\infty_{n=-\infty} (2 - 2 n k) = \sum^\infty_{n=-\infty} 
2 = 2+4\zeta_{R}(0) = 0 \ .
\ee
Thus,   as in  the 1-loop  correction 
in the  case of the  M2 brane wrapped on 
AdS$_{2}\times S^{1}$ 
   in \ci{Giombi:2023vzu},
the UV divergences  coming  from the tower of $n\neq 0$   modes  effectively cancel the  non-zero   contribution of the $n=0$ string modes  so that  the full   M2 brane 
1-loop correction is  UV finite. 
As discussed in  \ci{Giombi:2023vzu},   this cancellation can be understood as being a consequence of the  fact 
 that  the 2d model   with  an  infinite set of modes  considered 
  here  is equivalent  to the original  3d  model  where there  are no logarithmic UV  divergences.\foot{To be precise, the relation between 2d and 3d UV divergences requires the use of an analytic regularization that  disposes of all power divergences and 
  is effectively consistent with symmetries  of the 3d theory that are restored in the UV limit.} 

  The vanishing  of $
 \zeta_{\rm tot}(0)$ implies also  that $\Gamma$ in \rf{777} 
 will not depend on the radius $L$ of the 2-sphere so 
 that the result  will only be a  function  of the  dimensionless  parameter $k$  in the spectrum. 

Let us first  consider  the part of the sum in \rf{777}   which  comes  from 
 the $n=0$  (string-level) modes.  The contribution of the $n=0$ modes in Tables 1 and 2    can be computed  using 
the expressions for the corresponding determinants in \rf{5.4},\rf{5.7},\rf{op},\rf{5.21}
and is found to vanish
\be \la{639}
\Gamma_{0} = 2\,\Big[s_{\frac{1}{2}}(\tfrac{1}{4}) + s_{1}(1)\Big]
 -2\,\Big[\wt s_{\frac{1}{2}}(\tfrac{1}{4},0)+\wt s_{0}(1; 1)\Big]=0 \ .\ee  
To show the  vanishing we used 
  the relation \rf{666}  between $s_p$ and $\wt s_p$  and dropped  all 
  0-mode  contributions (corresponding to the lowest values of $\ell$ in the 
  respective sums)  that should be treated separately.
That all   the non-trivial finite   contributions  of the $n=0$ string-theory  modes 
mutually  cancel was   observed already  in  \ci{Gautason:2023igo}.

Next, let   us    determine  the contributions of the  rest of the  M2 brane fluctuation modes  with $n\neq 0$. 
Using the data in Table 1  and  the expression in \rf{5.21}  we find that 
 the bosonic mode contribution to $\Gamma$ in \rf{777} is
\ba\la{63}
\Gamma_{B} &= 2\sum_{n\neq 0} \Big[s_{\frac{1+|1+nk|}{2}}(1)+s_{\frac{1+|n|k}{2}}(\tfrac{1}{4})\Big] = 2\sum_{n=1}^{\infty}\Big[
s_{\frac{nk}{2}+1}(1)+2s_{\frac{nk}{2}+\frac{1}{2}}(\tfrac{1}{4})+
s_{\frac{nk}{2}}(1)\Big].
\ea
Similarly, in  the  fermionic case  we are to use the expression in \rf{5.7},\rf{op} 
where $\wt s_p(\mu)$ is defined  in terms of  the finite 
expression for $s_p(\mu)$  in \rf{5.21}. Separating the $n>0$ and $n<0$ contributions\foot{Note 
 that the  values  of ${\sfm }$ in Table \ref{tab:fer} depend  on the sign of $n$.}
 we  obtain 
\ba
\Gamma_{F} &=  
\sum_{n=1}^{\infty} \Big[\wt s_{\frac{nk+1}{2}}(\tfrac{1}{4}; \tfrac{nk}{2})+\wt s_{\frac{nk-1}{2}}(\tfrac{1}{4}; \tfrac{nk}{2}-1)+2\,\wt s_{\frac{nk}{2}}(1; \tfrac{nk}{2}+1)\Big]\lp \ \ 
+\sum_{n=1}^{\infty} \Big[\wt s_{\frac{nk-1}{2}}(\tfrac{1}{4}; \tfrac{|n|k}{2}-1)+\wt s_{\frac{nk+1}{2}}(\tfrac{1}{4}; \tfrac{nk}{2})+2\,\wt s_{\frac{nk}{2}}(1; \tfrac{nk}{2}+1)\Big]\lp
= 2\sum_{n=1}^{\infty} \Big[\wt s_{\frac{nk+1}{2}}(\tfrac{1}{4}; \tfrac{nk}{2})+\wt s_{\frac{nk-1}{2}}(\tfrac{1}{4}; \tfrac{nk}{2}-1)+2\,\wt s_{\frac{nk}{2}}(1; \tfrac{nk}{2}+1)\Big].\la{64}
\ea
If we now use the  definition of  $\wt s$ in  (\ref{5.7}),
i.e. $\wt s_{p}(\mu; {\sfm }) \equiv   2 p\,\log{\sfm }+s_{p+1}(\mu)$,   we find 
\ba
\Gamma_{F} =& 2\sum_{n=1}^{\infty} \Big[
(nk+1)\log\tfrac{nk}{2}+s_{\frac{nk+3}{2}}(\tfrac{1}{4})
+(nk-1)\log(\tfrac{nk}{2}-1)+s_{\frac{nk+1}{2}}(\tfrac{1}{4})\lp\qquad \ \ 
+2\,nk\log(\tfrac{nk}{2}+1)+2\, s_{\frac{nk}{2}+1}(1)\Big]\la{65}\ .\ea
Hence,  combining \rf{63} and \rf{65} we get 
\ba 
\Gamma= \Gamma_B -\Gamma_F =& 2\sum_{n=1}^{\infty}\Big[
s_{\frac{nk}{2}}(1)-s_{\frac{nk}{2}+1}(1) +s_{\frac{nk}{2}+\frac{1}{2}}(\tfrac{1}{4}) -s_{\frac{nk}{2}+\frac{3}{2}}(\tfrac{1}{4})
\lp\ \ \ \ \ \ \ \ \ 
-(nk+1)\log\tfrac{nk}{2}
-(nk-1)\log(\tfrac{nk}{2}-1)
-2\,nk\log(\tfrac{nk}{2}+1)\Big].\la{668}
\ea
Here the  first line can be simplified  
with the help of  the relation for $s_{p+1}$  used in \rf{666} (cf \rf{5.3}), i.e. 
$
 s_{p+1}(\mu)-  s_{p}(\mu)  = -2p \log(p^{2}-\mu)$.
 Then   we observe  a   remarkable  cancellation 
 (which  should  be  attributed again  to an underlying supersymmetry)
of  all   the  non-trivial  functions $s_{p}(\mu)$ (cf. \rf{5.21})
with only  few  logarithms remaining.  

Thus, notwithstanding the complexity of the intermediate expressions, the 
  final result  for $\Gamma$  is   very   simple 
\be
\la{6.9}
\Gamma = 2\sum_{n=1}^{\infty}\log\Big(\tfrac{n^{2}k^{2}}{4}-1\Big).
\ee
Surprisingly, this is the same  expression (up to overall factor of 2) 
as  was found in \cite{Giombi:2023vzu}  for the 1-loop correction  
in the case of the AdS$_{2}\times S^{1}$   M2 brane solution, 
despite the fact that the two  fluctuation spectra are   very different. 
This is, however, in line with the fact that the prefactors in the localization results in \rf{z} and \rf{zz}
happen to be  closely related.

The sum in \rf{6.9}  may  thus be  computed in the same way as in \cite{Giombi:2023vzu}
(i.e. using again the Riemann $\zeta$-function regularization, namely,
 $\zeta_{R}(0)=-\ha, \  \zeta'_{R}(0)=-\ha \log (2 \pi)$): 
\ba
\Gamma & = 4\sum_{n=1}^{\infty}\log\tfrac{nk}{2}+2\,\sum_{n=1}^{\infty}\log\Big(1-\tfrac{4}{n^{2}k^{2}}\Big)\no \\
&=4\log\tfrac{k}{2}\, \zeta_{R}(0) -  4 \zeta'_{R}(0)
+ 2  \log\Big(\frac{k}{2\pi}\sin\frac{2\pi}{k}\Big)
= 2\, \log\Big(2\, \sin\frac{2\pi}{k}\Big)\ . \la{610}
\ea
It is interesting to note that \rf{6.9}   may   be interpreted as  twice 
a  1d ``massive'' determinant on $S^1$ or    the   contribution of 
a loop of a   point 
 particle in the  (inverted)  harmonic oscillator potential 
 on a   circle ($s\equiv s+2 \pi$)\foot{Note
  that the Riemann $\zeta$-function regularization used in \rf{610} 
is the  standard prescription of how to  define similar 1d determinants appearing  in various path integral  representations.} 
\be  \la{00}
\Gamma= 2 \times  \ha \log {\rm det}' \big({\te -  {k^2\ov 4} {d^2\ov d s^2}  }-1\big)
=2 \sum_{n=1}^{\infty}\log\Big(\tfrac{k^{2}}{4}n^2 -1\Big)\ , 
\ee
where prime means   we project out the negative  mode  corresponding to $n=0$
(there is no zero modes if $k>2$). 
Thus the  simple form of the final  result \rf{6.9}  suggests that 
  what happens is that out of  all the fluctuation modes of the M2 brane in the instanton background only 
two   particular  bosonic 1d modes  survive after the fermion-boson cancellations,
reminiscent of some kind of  localization. 
The same reduction to just one  1d  bosonic mode happened in the case of the AdS$_2 \times S^1$ membrane 
computation in \cite{Giombi:2023vzu}. 

We   conclude  that the 1-loop instanton prefactor in \rf{413}  is 
\be \la{160}
{\rm Z_1} =  \Ntot \,  e^{-\Gamma} = \Ntot \, {1 \ov 4 \sin^2 (\frac{2\pi}{k}) } 
\ ,  \ee
where we introduced   a numerical ($k$-independent) factor  $\Ntot$  
to account for  the contribution  of  
the 0-modes that we omitted above
and  also of possible  degeneracy of the instanton saddle  contributions.  
In the large $k$ limit this reduces to  (cf.  \rf{w}) \ 
${\rm Z_1}= \Ntot {k^2 \ov 16 \pi^2} + ...=     {2\Ntot\ov \pi} {T\ov  \gs^2} + ...$, i.e. to 
 the expression found in \cite{Gautason:2023igo}.
 Here we get  it  directly as  a limit  of  the UV finite M2  brane  contribution, 
 without need  to  
fix   the form of the overall  factor  by some  indirect considerations
 as was   done in \cite{Gautason:2023igo}
using an analogy with 
 the string Wilson loop case  computation in \ci{Giombi:2020mhz}.

 To determine   the value of  $\Ntot$ let us note first  that  we are  to  add a  factor of  2  due to
the equal  instanton   and anti-instanton contributions 
(these become  distinguished  if there is a constant $C_3$   background). 
 We also need a further factor of 2   that  was argued in  \cite{Gautason:2023igo}
 to represent  the contribution of the  0-modes of the string fluctuations.\foot{The contributions 
 of the equal number of 12+12  bosonic and fermionic modes  can  be regularized 
 and shown to  cancel   after introducing a squashing parameter in the  $\CP^3$ 
metric. Then there are     two   possible string $\CP^1$ saddles  contributing equally 
 and thus  giving an extra  factor of 2   \cite{Gautason:2023igo}.}
It would be very  interesting to derive this result systematically 
by  introducing the collective  coordinates for the bosonic and fermionic 
0-modes and  computing the volume of the corresponding supercoset   
(that may turn out to be finite like the volume of the superM\"obius  group on the disk \ci{Andreev:1988cb}).\foot{Let us 
also recall  that the equal number of the bosonic and fermionic  0-modes in the $\CP^1$ instanton background are found 
also in the (2,2) supersymmetric 2d $\CP^1$  sigma model \ci{Novikov:1984ac}  where their  contributions
to the instanton measure  (and thus e.g. to the $\beta$-function) effectively cancel each other.}

To conclude, using that $\Ntot=4$  
 and accounting for the minus sign in \rf{sss},
we  precisely match  the 1-instanton  prefactor  in the localization  result in \rf{z}.

\section{Concluding remarks and open problems}

In this paper we  presented a new remarkable  test of  the 
AdS$_4$/CFT$_3$ duality between 
ABJM theory with large rank of the gauge group 
$N$ and   finite  level $k$   and M-theory   on  AdS$_4 \times S^7/\ZZ_k$. 
We  reproduced the leading (at large $N$ and fixed $k>2$)   instanton   prefactor in the 
localization result for  the non-perturbative   part of the ABJM free   energy on $S^3$ in \rf{z} 
from a  quantum 1-loop  correction to the classical action factor of the M2 brane $S^3/\ZZ_k$ instanton.  
This generalizes to finite $k$  the   analysis    of  the  string $\CP^1\subset \CP^3$  instanton contribution 
in type IIA string theory \ci{Cagnazzo:2009zh,Gautason:2023igo}. 

Like in  the earlier   AdS$_2\times S^1$  M2 brane   example   in \ci{Giombi:2023vzu},   the 
quantum 1-loop M2 brane computation  described above was   fully consistent (UV finite) 
and gave an unambiguous  prediction for the resulting function of $k$ in agreement with  \rf{z}.
One    subtle  issue  that  would  be interesting  to clarify further is  the  factor 2  associated 
with the  zero mode  contribution  that originates   from the string-level fluctuations; here we   fixed it  following 
 \ci{Gautason:2023igo}  but there may  exist  a more direct derivation. 

There are several possible  extensions of our work. 
One may consider the  leading perturbative $1\ov \sqrt N$   correction to the prefactor in \rf{z},\rf{2.11}
\be \la{733}
F^{\rm inst} (N,k) 
= - \frac{1}{\sin^{2} (\frac{2\pi}{k})}    \Big[1 + {1\ov \sqrt N}  h_1(k) + ... \Big] \, e^{-2\pi  \sqrt{2N\ov k }} + ...\ ,  
\qquad \ \   h_1(k) =\te {\pi \ov  \sqrt {2k}  }  {k^2-40\ov 12k } \ ,
\ee 
and try  to reproduce the coefficient  $h_1(k)$ from the 2-loop   M2 brane  correction, 
which  should come  with a factor of  the inverse of the  effective M2 brane tension in \rf{310}, i.e. 
${\rm T_2}^{-1}  =  {\pi \ov \sqrt{ 2 k} } {1\ov \sqrt N}$.\foot{As was suggested in \ci{Giombi:2023vzu},  a  
similar 
2-loop  computation  in the case of the   AdS$_2\times S^1$   M2 brane 
surface should reproduce the 
coefficient of the $1\ov \sqrt N$ correction to   the prefactor of  the Wilson loop 
expectation value   in  \rf{zz}. Note that  the  string theory values of the coefficients of these  $1\ov \sqrt N$  corrections 
 in \rf{733} and \rf{zz}
are   sensitive to the  precise  form of the relation  between  the string theory parameters in \rf{310}
 and   gauge theory  parameters  $N,k$, i.e. to the shift 
 $ N \to N - {1\ov 24} (k-k^{-1})$ suggested in  \cite{Bergman:2009zh}. } 
Such a 2-loop calculation would require the  use of  the quartic bosonic and fermionic terms in the corresponding supermembrane  action \ci{deWit:1998yu,Pasti:1998tc,Claus:1998fh}.
 It   would be important  to check if the  2-loop M2  brane contribution is, in fact, 
  UV finite,   despite the apparent  non-renormalizability of the   membrane action.\foot{Such  2-loop  computation 
  is, however,   going to be    hard as the  background 3d  metric in the static gauge \rf{47}  is  that of 
  $S^3/\ZZ_k$, i.e.   is  not flat.}

As was   mentioned  in  section  2, the  subleading (at large $N$ and $k>2$) 
 non-perturbative  contributions 
to the ABJM free energy  involve,  in addition to   the contributions of M2  brane instantons 
wrapping $S^3/\ZZ_k \subset  S^7/\ZZ_k$,    also the contributions  of M2 brane   instanton
 wrapping  $\RP^3$  3-cycle  in $\CP^3\subset S^7/\ZZ_k$  (not involving 11d circle),
 corresponding to the  D2-brane  instanton  in the type IIA string  limit. 
 One may   wonder  if  the analog of the  semiclassical M2  brane  computation  that we performed 
  in this paper 
 for the  first  ($S^3/\ZZ_k$)  instanton can be  also repeated  for the second  ($\RP^3=S^3/\ZZ_2$) one,  
 thus  determining   a $k$-dependent 
 prefactor of  the corresponding exponential  $ e^{- \pi \sqrt {2Nk}  }$  in \rf{2.6}. 
 
 As we 
 discuss  in Appendix \ref{adD2}, 
  in this second case the M2 brane solution 
  is not wrapping the  11d circle in \rf{3.1}  and thus  the corresponding induced 3-metric and
 the quantum M2 brane  fluctuation determinants do not  have a non-trivial dependence   on $k$  (cf. \rf{3.8},\rf{4.4}).  
At the same time, the localization  result for  the prefactor of  $ e^{- \pi \sqrt {2Nk}  }$  in \rf{2.6} 
  vanishes for odd $k$  has  singular  dependence on  even  $k$  (see  \ref{74}).
  We point out that  for even $k$   the contribution of the $\RP^3$   instanton 
  in fact  ``mixes''   with that of $k\ov 2$-instanton  of the $S^3/\ZZ_k$ type 
  (both have the same classical action)   and thus    should not be considered in isolation. 
 In general, one is  to combine together 
  all of  the  contributions  in \rf{2.6}  that   have the same exponential factor  and then 
  the prefactor   will have a regular value for  any  given integer $k$  \cite{Calvo:2012du}.

  In particular,  one may study the special  cases of $k=1,2$. 
  While  for 
   $k\ge 2$  the
 leading non-perturbative correction to $F$ in \rf{2.6} is given by 
the single $(n_{\tc}, n_{\ttc}) = (1,0)$ instanton  contribution, 
 for $k=1,2$  the two terms in the exponent in \rf{2.6} become of  the same order
suggesting that  the two types of instantons  should be treated on an equal footing. 
We comment  on this further  in  Appendix \ref{adD2}.

In the  1-loop quantum M2  brane   computation near the $S^3/\ZZ_k$ instanton 
described in this paper  we assumed $k>2$   but   it  is  
straightforward  to extend it  to $k=1,2$    by   determining  the fluctuation 
spectra directly in these  special cases (like that was done  in  the AdS$_2\times S^1$    case in 
\ci{Giombi:2023vzu}). In particular, for $k=1$ we find  that  there are 4 negative  modes 
(corresponding to $n=-1$ for each of the 4  $\CP^3$ fluctuations in Table 1)
 reflecting instability of the $S^3$ instanton in $S^7$  in line with $\pi_3(S^7)$ being trivial. 
 In addition,  there are   extra  0-mode contributions 
that  appear in these  cases   with enhanced supersymmetry.

  Another  interesting open problem  is to try to generalize 
  both 
  the computation in \ci{Giombi:2023vzu}
  and in the present paper 
  in order to determine  the  leading   instanton correction  to the expectation 
  value  of the   circular BPS  Wilson loop  from  the quantum M2  brane theory
  to  match the corresponding localization result 
  on the ABJM gauge theory side   \cite{Hatsuda:2013yua} (cf.  \ci{Okuyama:2016deu}).\foot{
   The 
    leading  instanton corrections to the $\frac{1}{2}$-BPS Wilson loop  in the fundamental representation
    is given by  \cite{Hatsuda:2013yua}
\ba
\langle W\rangle = 
 \int_{-i\infty}^{i\infty}\frac{d\mu}{2\pi i} \ e^{-N\mu}\  
 \frac{e^{\frac{2\mu}{k}}}{2\,\sin\frac{2\pi}{k}} \  \Q (\mu, k), 
  \qquad 
\Q = 1+2Q+3Q^{2}+10Q^{3}+\Big(49-32\sin^{2}\frac{2\pi}{k}\Big)\, Q^{4}+\cdots, \ \  Q \equiv  -e^{-\frac{4\mu}{k}} .\no 
\ea
 Here  the corrections due to 1,2,3 instantons do not introduce a new  $k$-dependence. $\Q$  has a smooth $k\to\infty$ limit 
 which is known in exact form 
from the instanton corrections to the disk amplitude in topological string  model \cite{Aganagic:2001nx}.
\iffa \be
(Q/z)^{1/2} = 1+2Q+3Q^{2}+10Q^{3}+49Q^{4}+\cdots,
\ee
with $z=z(Q)$ given by the solution to 
\be
\frac{1}{2}\log\frac{Q}{z} = 2\,z {}_{4}F_{3}(1,1,\tfrac{3}{2}, \tfrac{3}{2}; 2, 2, 2; 16 z),
\ee
to be solved for $z\to 0$ (and then $Q=z+4z^{2}+\dots$).\fi 
}
  The corresponding minimal 3-surface  should  be   a  superposition of 
  AdS$_2\times S^1$ in  \ci{Sakaguchi:2010dg,Giombi:2023vzu}   and $S^3/\ZZ_k$  discussed  above. 
 Here  a natural   static gauge choice  may be  along the AdS$_2\times S^1$ subspace 
 as in  \ci{Giombi:2023vzu},  
 implying a more involved  structure of fluctuations  in all of  $\CP^3$ directions.\foot{In particular, one    would   need to 
 include explicitly  the contribution of the ``longitudinal''  fluctuations along the $\CP^1\subset \CP^3$ directions.}

\section*{Acknowledgements}
We  are grateful to  F. Gautason,  V.G.M. Puletti and J. van Muiden    for  useful
  communications, 
  sharing some unpublished details  of  their  work  \ci{Gautason:2023igo}      and  comments on the draft. 
We also thank 
M. Marino, S. Pufu, R. Roiban,  L. Wulff  and K. Zarembo
for useful  communications   and discussions
and N. Drukker for important  comments on the draft. 
MB was supported by the INFN grant GSS (Gauge Theories, Strings and Supergravity). 
SG is supported in part by the US NSF under Grant No.~PHY-2209997.
AAT is supported by the STFC grant ST/T000791/1. He also acknowledges
the hospitality of  Nordita at the final stage of this work. 

\appendix
\section{
Comments on M2 brane   instantons wrapping $\RP^3 \subset \CP^3$ of $S^7/\ZZ_k$} \la{adD2}

To recall, 
the non-perturbative corrections to the free energy in (\ref{2.6})
are labeled by a pair of integers $(n_{\tc}, n_{\ttc})$ 
with $(n_{\tc},0)$ and $(0,n_{\ttc})$   
representing the two types   ($S^3/\ZZ_k$ and $\RP^3$)  of (multi)instantons.  
 The corresponding  terms   in the  non-perturbative    part of 
 the grand potential \rf{2.2} are   (cf.  (\ref{y})
\ba
J^{\rm np}(\mu,k) = & \sum_{n_{\tc},n_{\ttc}=0}^{\infty}J^{\rm np}_{(n_{\tc}, n_{\ttc})}(\mu,k)\ ,\la{71}
\\
J^{\rm np}_{(n_{\tc},0)} = d_{n_{\tc}}(k)\, e^{-\frac{4n_{\tc}}{k}\mu}\ , &\qquad\qquad 
J^{\rm np}_{(0,n_{\ttc})} = \big[a_{n_{\ttc}}(k)\,\mu^{2}+b_{n_{\ttc}}(k)\, \mu+c_{\ttc}(k)\big]\, e^{-2n_{\ttc}\mu}\ .\la{72}
\ea
These  translate into the coefficients appearing  in the prefactor of the  non-perturbative part  of the 
free energy \rf{2.6}. 
In particular, the leading $(n_{\tc}, n_{\ttc})=(0,1)$ contribution   to $F^{\rm inst}$ in (\ref{213})   is \cite{Hatsuda:2012dt}
(cf.  the $(1,0)$  contribution  in  (\ref{2.11}))
\ba
\la{x73}
F^{\rm inst}_{(0,1)}(k,N)=& -\Big[C(k)^{-1}\big(N+2-B(k)\big)\, a_{1}(k)-c_{1}(k)\Big]\,\frac{\text{Ai}[C(k)^{-{1\ov3}}(N+2-B(k))]}{\text{Ai}[C(k)^{-{1\ov 3}}(N-B(k))]}\lp
-C(k)^{-{1\ov 3} }\, b_{1}(k)\,\frac{\text{Ai}'[C(k)^{-{1\ov 3}}(N+2-B(k))]}{\text{Ai}[C(k)^{-\third}(N-B(k))]}.
\ea
The  coefficients $d_{n_{\tc}}(k)$  corresponding to  multi-instantons of the first  type 
 may be computed  using  the  topological string  representation   \cite{Hatsuda:2012dt}
(cf. \rf{yy}) 
\be
d_{1}(k) = \frac{1}{\sin^{2}\frac{2\pi}{k}}, \qquad 
d_{2}(k) = -\frac{1}{2\,\sin^{2}\frac{4\pi}{k}}-\frac{1}{\sin^{2}\frac{2\pi}{k}}, \qquad 
d_{3}(k) = \frac{1}{3\,\sin^{2}\frac{6\pi}{k}}+\frac{3}{\sin^{2}\frac{2\pi}{k}}, \quad \cdots \ . \la{73}
\ee
The coefficients  $a_1, b_1, c_1$  corresponding to  the 
 single   instanton of the second kind in \rf{72}  have been conjectured in \cite{Hatsuda:2012dt,Hatsuda:2012hm,Calvo:2012du} and obtained systematically from the refined topological string
 representation  in \cite{Hatsuda:2013oxa}
\ba
&a_{1}(k) = -\frac{4}{\pi^{2}k}\cos\frac{\pi k}{2}, \qquad \qquad b_{1}(k) = \frac{2}{\pi\tan\frac{\pi k}{2}}\cos\frac{\pi k}{2}, \quad\no  \\
&c_{1}(k) = \Big(
-\frac{2}{3k}+\frac{5k}{12}+\frac{k}{2\sin^{2}\frac{\pi k}{2}}+\frac{1}{\pi\tan\frac{\pi k}{2}}\Big)\cos\frac{\pi k}{2}.\la{74}
\ea
Note that these vanish for odd $k$ and two of them are singular for  even $k$. 

In the dual  M-theory  setting, 
the M2  brane   wrapped on  $\RP^3$ 3-cycle in $\CP^3$ which is part of $ S^7/\ZZ_k$   with 
 the metric  \rf{3.1}  can be explicitly described as follows   \ci{Drukker:2011zy}. 
Using the    angular   parametrization of $\CP^{3}$ (here $\alpha, \theta_{1}, \theta_{2}\in [0,\pi]; \ \chi, \varphi_{1},\varphi_{2}\in[0,2\pi]$)\foot{In this parametrization 
$A$  in \rf{3.1}  is given by $A = \frac{1}{2}\big(\cos\alpha\,  d\chi+\cos^{2}\frac{\alpha}{2}\,\cos\theta_{1}d\varphi_{1}+\sin^{2}\frac{\alpha}{2}\,\cos\theta_{2}d\varphi_{2}\big).
$}
\ba
ds^{2}_{\CP^{3}} =&\frac{1}{4}d\alpha^{2}+\frac{1}{4}\cos^{2}\frac{\alpha}{2}(d\theta_{1}^{2}+\sin^{2}\theta_{1}d\varphi_{1}^{2})
+\frac{1}{4}\sin^{2}\frac{\alpha}{2}(d\theta_{2}^{2}+\sin^{2}\theta_{2}d\varphi_{2}^{2})\lp
+\frac{1}{4}\sin^{2}\frac{\alpha}{2}\cos^{2}\frac{\alpha}{2}\,(2d\chi+\cos\theta_{1}d\varphi_{1}-\cos\theta_{2}d\varphi_{2})^{2}, 
\ea
the  relevant classical  M2 brane solution  is wrapped on 
 $\mathbb{RP}^{3}\subset \CP^3 $  parametrised  by  $\theta_{1}=\theta_{2}, \, \varphi_1=-\varphi_2,\, \chi$   and $\alpha=0$ \ci{Drukker:2011zy,Park:2020hgt}.
 The static gauge adapted to this solution is thus 
  (here we  label  the world-volume coordinates as $\xi^i= (s_1,s_2,s_3)$, cf. \rf{45})  
\be
\theta_{1}=\theta_{2}=s_1\in [0,\pi], \qquad   \varphi_{1}=-\varphi_{2}=s_2\in[0,2\pi] , \qquad  \chi=s_3\in[0,2\pi] \ . \la{86}
\ee 
The   induced world-volume metric  is that of $\RP^3$, i.e. 
$ds^2 = \frac{R^{2}}{4}  [  ds_1^2 + \sin^2 s_1  ds_2^2  + (ds_3 + \cos s_1 ds_2)^2]$  which  is 
the standard  metric of $S^3$ of  radius $R$ in Hopf parametrization  but with the angle  $s_3$ having period $2 \pi$ instead of   $4\pi$:
\be
g_{ij} 
= \frac{R^{2}}{4} \begin{pmatrix}
1 & 0 & 0 \\
 0 & 1 & \cos s_1 \\
 0 & \cos s_1 & 1
\end{pmatrix}, \qquad\ \ \ \  \sqrt{g} = \frac{R^{3}}{8}\sin s_{1} \ . \la{87}
\ee
 The  classical  value of the M2 brane action  is then given by  (cf. \rf{410},\rf{3.12})
\be
\IS_{\rm cl} = T_{2}\frac{R^{3}}{8} (2\pi)^2 \int_{0}^{\pi} ds_1 \, \sin s_{1} 
 = \pi\, \sqrt{2kN} \ . \la{88}
\ee
This   is of course the same as   the action of the corresponding D2 brane wrapped on $\RP^3 \subset \CP^3$  in type IIA theory 
and  is also   the value  appearing as  a factor of  $n_{\ttc}$ in the exponent in   (\ref{2.6}).

Expanding the  M2  brane action \rf{m}   near this solution  we observe that, as the 11d angle $y$ in \rf{3.1} 
has a trivial classical value,
  the quadratic  fluctuation  Lagrangian  will depend on $k$ only via  the 
 term $ {1\ov k^2} (\del \td y)^2$.\foot{The classical value of  the 1-form  $A$ in \rf{3.1} is 
$A= \ha ( ds_3 + \cos s_1 ds_2)$ but there are similar off-diagonal terms in \rf{87} 
so the  term ${1\ov k}  \del \td y  A$   will  also not lead to  a non-trivial contribution. 
Note  also that the leading quadratic part of the fermionic Lagrangian  can not depend on $k$ as the bosonic 
coordinates have classical values only along  the $\RP^3 \subset \CP^3$.
 }
 Thus, in contrast to what we  found 
 for the $S^3/\ZZ_k$   instanton,  here  the corresponding 1-loop M2  brane  partition function
 will not have a non-trivial  dependence on $k$.\foot{Rescaling of $\td y$   by $k$ will not  lead to 
 a  factor  of $k$ in the partition function 
 as the analog of $\zeta(0)$   vanishes in 3d.}

At the same time, the localization  coefficients \rf{74}  have a peculiar dependence on $k$. 
They do not appear to admit a regular large $k$ limit,  vanish for odd  values of $k$  and are singular for even $k$.
However, it is important  to realize  that to compare the localization   result  to   M2  brane   partition  function 
we need  to add together  the prefactors of all   terms in \rf{2.6}  that  have  the same  value of  the exponentials, i.e. 
$\exp\big[-2\pi\sqrt{N}\, \big(n_{\tc} \sqrt {2\ov  k}+n_{\ttc}  \sqrt{k\ov  2}\, \big)\big] $.

For  $k$   even  that means that $({n_{\tc}, n_{\ttc}})$= $({k\ov2},0)$ and $(0,1)$  terms 
 should be treated on an equal footing.  
 Indeed,  
 the corresponding $k\ov 2$-instanton of the $S^3/\ZZ_k$ type  has the same action   as  the 
 1-instanton of the $\RP^3 =S^3/\ZZ_2$ type. 
 This  implies  that  the $\RP^3$  instanton should not be considered in isolation, 
 resolving the  above puzzle  about  $k$-independence of the  corresponding 1-loop  M2   contribution:
 it is  the combined 1-loop M2 brane partition functions  of the  $({k\ov2},0)$ and $(0,1)$   instantons  that should  
 be then compared to localization. 
 If $k$ is  odd,   the same remark  should  apply  to the  contribution of the 
  $({k},0)$ and $(0,2)$   instantons  (which is of course subleading to the contributions of 
  $({k-1\ov 2} ,0)$ and $(0,1)$   instantons).



In the main part of this  paper we have been assuming  $k > 2$ in which case the
 leading non-perturbative correction to $F$ in \rf{2.6} is given by 
the single $(n_{\tc}, n_{\ttc}) = (1,0)$ instanton  contribution. 
 It is, however,  necessary to include the contributions of the second  type of M2  brane 
  instanton in order  to  compute non-perturbative  corrections to the  free energy  
for  the lowest values of $k=1,2$. 
These values are special since the  corresponding ABJM model
 (and also  its dual M-theory counterpart)  should have the  enhanced supersymmetry: 
from $\mc N=6$ to $\mc N=8$ \cite{Aharony:2008ug,Bagger:2012jb,Halyo:1998pn,Gustavsson:2009pm}.

Indeed, for $k=1,2$  the two terms in the exponent in \rf{2.6} become of  the same order
and should be treated on an equal footing. 
The relevant combinations of  the two instanton numbers are\footnote{For $k=1$ the dominant contribution 
would naively come  from  the  $(0,1)$ contribution 
but this term is actually absent \cite{Putrov:2012zi}; it is  conjectured to vanish for all odd $k$, and that was  checked for $k=5,7$ in \cite{Hatsuda:2012hm}. Note that this is consistent with the expressions in (\ref{74}), which vanish for odd $k$.} 
\iffa \begin{alignat*}{2}
 &k=1:\qquad     & & (1,0)+(0,2),    \\
 &k=2:   & & (1,0)+(0,1).
\end{alignat*}\fi
\be \la{345}
k=1:\qquad   (1,0)+(0,2) \ ; \qquad \qquad\qquad \qquad  k=2:   \qquad  (1,0)+(0,1)  \ . \ee 
For $k=1$  we thus have the  $S^3$ one-instanton    and  the    $\RP^3= S^3/\ZZ_2$    two-instanton  
which are indeed essentially the same. Similarly, for $k=2$ both the $(1,0)$ and $(0,1)$ instanton correspond 
to $S^3/\ZZ_2$. 

\iffa the 
 non-perturbative contributions   in \rf{2.6}   coming  from $({n_{\tc}, n_{\ttc}})$= $({k\ov2},0)$ and $(0,1)$ 
 have the same exponential weight  and thus   should be treated on an equal footing 
 (the corresponding $k\ov 2$-instanton of the $S^3/\ZZ_k$ type  has the same action   as  the 
 1-instanton of the $\RP^3 =S^3/\ZZ_2$ type). 
\fi

Note  that for these values  of $(n_{\tc}, n_{\ttc})$ the  corresponding  coefficients in \rf{72},\rf{73},\rf{74} are singular 
but their sums are   finite and give  the following expressions for  $J^{\rm np}$ in \rf{71} 
\ci{Hatsuda:2012hm} \footnote{
Let us note also that for generic even $k$ 
the sum of the $({n_{\tc}, n_{\ttc}})$= $({k\ov2},0)$ and $(0,1)$ 
  contributions   is again finite 
after cancellations between the two separately divergent  terms \cite{Calvo:2012du}.}
\be
J^{\rm np}(\mu, 1) = \frac{1}{4\pi^{2}}(16\mu^{2}+4\mu+1)e^{-4\mu}+\cdots ,\qquad 
J^{\rm np}(\mu, 2) =  \frac{1}{\pi^{2}}(4\mu^{2}+2\mu+1)e^{-2\mu}+\cdots.\la{710}
\ee
These expressions for the grand potential   plugged into \rf{2.1} give 
 the following leading non-perturbative contributions to $F^{\rm inst}$ in (\ref{213})  for $k=1$ and $k=2$ 
 \cite{Hatsuda:2012dt}
\ba
F^{\rm inst}_{\rm leading}(1, N) &= -\Big(2N+\frac{C(1)}{8}+\frac{29}{4}\Big)\,\frac{\text{Ai}[C(1)^{-\third}(N+\frac{29}{8})]}{\text{Ai}[C(1)^{-\third}(N-\frac{3}{8})]}+\frac{C(1)^{2\ov 3}}{2}\,
\frac{\text{Ai}'[C(1)^{-\third}(N+\frac{29}{8})]}{\text{Ai}[C(1)^{-\third}(N-\frac{3}{8})]},\no  \\
F^{\rm inst}_{\rm leading}(2, N) &= -\big(4N+C(2)+7\big)\,\frac{\text{Ai}[C(2)^{-\third}(N+\frac{7}{4})]}{\text{Ai}[C(2)^{-\third}(N-\frac{1}{4})]}+2\,C(2)^{2\ov 3}\,
\frac{\text{Ai}'[C(2)^{-\third}(N+\frac{7}{4})]}{\text{Ai}[C(2)^{-\third}(N-\frac{1}{4})]}.\la{k12}
\ea
Expanded to  leading order in large $N$, these expressions  give, respectively, 
 for $k=1$  \cite{Putrov:2012zi}  and $k=2$  
\be
F^{\rm inst}_{\rm leading}(1, N) = -2\,N\,e^{-2\,\pi\, \sqrt{2N}}+\cdots,  \qquad\qquad 
F^{\rm inst}_{\rm leading}(2, N) = -4\,N\,e^{-2\,\pi\, \sqrt{N}}+\cdots \ . \la{a13}
\ee
Here the  $N$ factor comes from the $\mu^{2}$ term with the  coefficient $a_{\ttc}(k)$ in (\ref{72})
(cf. the first term  in  
(\ref{x73}) 
 in the  $k=2$ case). 
It would be very interesting to reproduce these  contributions to  $F^{\rm np}$  in 
\rf{2.6} 
by  a   quantum M2  brane computation.\foot{It is in principle straightforward  to repeat the 1-loop  analysis described in  sections 3-6 above  for the  special cases of 
$k=1,2$. Let us mention only that  the masses in the 
 bosonic and fermionic charged sectors  in Tables 1  and 2  depend on $2nk+n^2 k^2$
and thus take the same values for $n$ and $-n-2/k$. Hence, there is an extra degeneracy 
in the spectrum when $k=1,2$.
 This should be related to the world-volume   analog of the enhanced  target space 
  supersymmetry in these cases.}
  Note that the fact that $F^{\rm inst}_{\rm leading}(1, N)$   is real does not contradict the 
  instability of the $S^3$ instanton in $S^7$  as there are 4 negative  bosonic modes contributing $i^4=1$.


\small 
\bibliography{BT-Biblio}

\providecommand{\href}[2]{#2}\begingroup\raggedright\begin{thebibliography}{10}

\bibitem{Aharony:2008ug}
O.~Aharony, O.~Bergman, D.~L. Jafferis and J.~Maldacena,
  \emph{{${\mathcal{N}}\!=6$ Superconformal Chern-Simons-Matter Theories,
  M2-Branes and Their Gravity Duals}},
  \href{https://doi.org/10.1088/1126-6708/2008/10/091}{\emph{JHEP} {\bfseries
  10} (2008) 091} [\href{https://arxiv.org/abs/0806.1218}{{\ttfamily
  0806.1218}}].

\bibitem{Pestun:2016zxk}
V.~Pestun et~al., \emph{{Localization techniques in quantum field theories}},
  \href{https://doi.org/10.1088/1751-8121/aa63c1}{\emph{J. Phys.} {\bfseries
  A50} (2017) 440301} [\href{https://arxiv.org/abs/1608.02952}{{\ttfamily
  1608.02952}}].

\bibitem{Giombi:2023vzu}
S.~Giombi and A.~A. Tseytlin, \emph{{Wilson Loops at Large N and the Quantum
  M2-Brane}}, \href{https://doi.org/10.1103/PhysRevLett.130.201601}{\emph{Phys.
  Rev. Lett.} {\bfseries 130} (2023) 201601}
  [\href{https://arxiv.org/abs/2303.15207}{{\ttfamily 2303.15207}}].

\bibitem{Sakaguchi:2010dg}
M.~Sakaguchi, H.~Shin and K.~Yoshida, \emph{{Semiclassical Analysis of M2-brane
  in $AdS_4 x S^7 / Z_k$}},
  \href{https://doi.org/10.1007/JHEP12(2010)012}{\emph{JHEP} {\bfseries 12}
  (2010) 012} [\href{https://arxiv.org/abs/1007.3354}{{\ttfamily 1007.3354}}].

\bibitem{Klemm:2012ii}
A.~Klemm, M.~Mari\~no, M.~Schiereck and M.~Soroush,
  \emph{{Aharony-Bergman-Jafferis--Maldacena Wilson Loops in the Fermi Gas
  Approach}}, \href{https://doi.org/10.5560/ZNA.2012-0118}{\emph{Z.
  Naturforsch. A} {\bfseries 68} (2013) 178}
  [\href{https://arxiv.org/abs/1207.0611}{{\ttfamily 1207.0611}}].

\bibitem{Drukker:2011zy}
N.~Drukker, M.~Marino and P.~Putrov, \emph{{Nonperturbative aspects of ABJM
  theory}}, \href{https://doi.org/10.1007/JHEP11(2011)141}{\emph{JHEP}
  {\bfseries 11} (2011) 141} [\href{https://arxiv.org/abs/1103.4844}{{\ttfamily
  1103.4844}}].

\bibitem{Hatsuda:2013gj}
Y.~Hatsuda, S.~Moriyama and K.~Okuyama, \emph{{Instanton Bound States in ABJM
  Theory}}, \href{https://doi.org/10.1007/JHEP05(2013)054}{\emph{JHEP}
  {\bfseries 05} (2013) 054} [\href{https://arxiv.org/abs/1301.5184}{{\ttfamily
  1301.5184}}].

\bibitem{Cagnazzo:2009zh}
A.~Cagnazzo, D.~Sorokin and L.~Wulff, \emph{{String instanton in AdS(4) x
  CP3}}, \href{https://doi.org/10.1007/JHEP05(2010)009}{\emph{JHEP} {\bfseries
  05} (2010) 009} [\href{https://arxiv.org/abs/0911.5228}{{\ttfamily
  0911.5228}}].

\bibitem{Gautason:2023igo}
F.~F. Gautason, V.~G.~M. Puletti and J.~van Muiden, \emph{{Quantized Strings
  and Instantons in Holography}},
  \href{https://arxiv.org/abs/2304.12340}{{\ttfamily 2304.12340}}.

\bibitem{Beccaria:2023hhi}
M.~Beccaria and A.~A. Tseytlin, \emph{{Comments on ABJM Free Energy on $S^3$ at
  Large $N$ and Perturbative Expansions in M-theory and String Theory}},
  \href{https://doi.org/10.1016/j.nuclphysb.2023.116286}{\emph{Nucl. Phys. B}
  {\bfseries 994} (2023) 116286}
  [\href{https://arxiv.org/abs/2306.02862}{{\ttfamily 2306.02862}}].

\bibitem{Marino:2011eh}
M.~Mari\~no and P.~Putrov, \emph{{ABJM Theory as a Fermi Gas}},
  \href{https://doi.org/10.1088/1742-5468/2012/03/P03001}{\emph{J. Stat. Mech.}
  {\bfseries 1203} (2012) P03001}
  [\href{https://arxiv.org/abs/1110.4066}{{\ttfamily 1110.4066}}].

\bibitem{Kapustin:2009kz}
A.~Kapustin, B.~Willett and I.~Yaakov, \emph{{Exact Results for Wilson Loops in
  Superconformal Chern-Simons Theories with Matter}},
  \href{https://doi.org/10.1007/JHEP03(2010)089}{\emph{JHEP} {\bfseries 03}
  (2010) 089} [\href{https://arxiv.org/abs/0909.4559}{{\ttfamily 0909.4559}}].

\bibitem{Marino:2009jd}
M.~Mari\~no and P.~Putrov, \emph{{Exact Results in ABJM Theory from Topological
  Strings}}, \href{https://doi.org/10.1007/JHEP06(2010)011}{\emph{JHEP}
  {\bfseries 06} (2010) 011} [\href{https://arxiv.org/abs/0912.3074}{{\ttfamily
  0912.3074}}].

\bibitem{Drukker:2010nc}
N.~Drukker, M.~Mari\~no and P.~Putrov, \emph{{From Weak to Strong Coupling in
  ABJM Theory}}, \href{https://doi.org/10.1007/s00220-011-1253-6}{\emph{Commun.
  Math. Phys.} {\bfseries 306} (2011) 511}
  [\href{https://arxiv.org/abs/1007.3837}{{\ttfamily 1007.3837}}].

\bibitem{Fuji:2011km}
H.~Fuji, S.~Hirano and S.~Moriyama, \emph{{Summing Up All Genus Free Energy of
  ABJM Matrix Model}},
  \href{https://doi.org/10.1007/JHEP08(2011)001}{\emph{JHEP} {\bfseries 08}
  (2011) 001} [\href{https://arxiv.org/abs/1106.4631}{{\ttfamily 1106.4631}}].

\bibitem{Hatsuda:2012dt}
Y.~Hatsuda, S.~Moriyama and K.~Okuyama, \emph{{Instanton Effects in ABJM Theory
  from Fermi Gas Approach}},
  \href{https://doi.org/10.1007/JHEP01(2013)158}{\emph{JHEP} {\bfseries 01}
  (2013) 158} [\href{https://arxiv.org/abs/1211.1251}{{\ttfamily 1211.1251}}].

\bibitem{Hanada:2012si}
M.~Hanada, M.~Honda, Y.~Honma, J.~Nishimura, S.~Shiba and Y.~Yoshida,
  \emph{{Numerical Studies of the ABJM Theory for Arbitrary $N$ at Arbitrary
  Coupling Constant}},
  \href{https://doi.org/10.1007/JHEP05(2012)121}{\emph{JHEP} {\bfseries 05}
  (2012) 121} [\href{https://arxiv.org/abs/1202.5300}{{\ttfamily 1202.5300}}].

\bibitem{Hatsuda:2014vsa}
Y.~Hatsuda and K.~Okuyama, \emph{{Probing Non-Perturbative Effects in
  M-theory}}, \href{https://doi.org/10.1007/JHEP10(2014)158}{\emph{JHEP}
  {\bfseries 10} (2014) 158} [\href{https://arxiv.org/abs/1407.3786}{{\ttfamily
  1407.3786}}].

\bibitem{Herzog:2010hf}
C.~P. Herzog, I.~R. Klebanov, S.~S. Pufu and T.~Tesileanu, \emph{{Multi-Matrix
  Models and Tri-Sasaki Einstein Spaces}},
  \href{https://doi.org/10.1103/PhysRevD.83.046001}{\emph{Phys. Rev. D}
  {\bfseries 83} (2011) 046001}
  [\href{https://arxiv.org/abs/1011.5487}{{\ttfamily 1011.5487}}].

\bibitem{Giombi:2020mhz}
S.~Giombi and A.~A. Tseytlin, \emph{{Strong coupling expansion of circular
  Wilson loops and string theories in AdS$_5 \times {\rm S}^5$ and AdS$_4
  \times {\rm CP}^3$}},
  \href{https://doi.org/10.1007/JHEP10(2020)130}{\emph{JHEP} {\bfseries 10}
  (2020) 130} [\href{https://arxiv.org/abs/2007.08512}{{\ttfamily
  2007.08512}}].

\bibitem{Bobev:2022eus}
N.~Bobev, J.~Hong and V.~Reys, \emph{{Large $N$ Partition Functions of the ABJM
  Theory}}, \href{https://doi.org/10.1007/JHEP02(2023)020}{\emph{JHEP}
  {\bfseries 02} (2023) 020}
  [\href{https://arxiv.org/abs/2210.09318}{{\ttfamily 2210.09318}}].

\bibitem{Aurilia:1980xj}
A.~Aurilia, H.~Nicolai and P.~K. Townsend, \emph{{Hidden Constants: The Theta
  Parameter of QCD and the Cosmological Constant of N=8 Supergravity}},
  \href{https://doi.org/10.1016/0550-3213(80)90466-6}{\emph{Nucl. Phys. B}
  {\bfseries 176} (1980) 509}.

\bibitem{Kurlyand:2022vzv}
S.~A. Kurlyand and A.~A. Tseytlin, \emph{{Type IIB supergravity action on M5 $
  \times $ X5 solutions}},
  \href{https://doi.org/10.1103/PhysRevD.106.086017}{\emph{Phys. Rev. D}
  {\bfseries 106} (2022) 086017}
  [\href{https://arxiv.org/abs/2206.14522}{{\ttfamily 2206.14522}}].

\bibitem{Aguilar-Gutierrez:2022kvk}
S.~E. Aguilar-Gutierrez, K.~Parmentier and T.~Van~Riet, \emph{{Towards an
  AdS$_{1}$/CFT$_{0}$ correspondence from the D({-}1)/D7 system?}},
  \href{https://doi.org/10.1007/JHEP09(2022)249}{\emph{JHEP} {\bfseries 09}
  (2022) 249} [\href{https://arxiv.org/abs/2207.13692}{{\ttfamily
  2207.13692}}].

\bibitem{Emparan:1999pm}
R.~Emparan, C.~V. Johnson and R.~C. Myers, \emph{{Surface terms as counterterms
  in the AdS / CFT correspondence}},
  \href{https://doi.org/10.1103/PhysRevD.60.104001}{\emph{Phys. Rev.}
  {\bfseries D60} (1999) 104001}
  [\href{https://arxiv.org/abs/hep-th/9903238}{{\ttfamily hep-th/9903238}}].

\bibitem{Duff:1987bx}
M.~J. Duff, P.~S. Howe, T.~Inami and K.~S. Stelle, \emph{{Superstrings in D=10
  from Supermembranes in D=11}},
  \href{https://doi.org/10.1016/0370-2693(87)91323-2}{\emph{Phys. Lett. B}
  {\bfseries 191} (1987) 70}.

\bibitem{Achucarro:1989dd}
A.~Achucarro, P.~Kapusta and K.~S. Stelle, \emph{{Strings From Membranes: The
  Origin of Conformal Invariance}},
  \href{https://doi.org/10.1016/0370-2693(89)90747-8}{\emph{Phys. Lett. B}
  {\bfseries 232} (1989) 302}.

\bibitem{Berman:2006vg}
D.~S. Berman and M.~J. Perry, \emph{{M-theory and the string genus expansion}},
  \href{https://doi.org/10.1016/j.physletb.2006.02.038}{\emph{Phys. Lett. B}
  {\bfseries 635} (2006) 131}
  [\href{https://arxiv.org/abs/hep-th/0601141}{{\ttfamily hep-th/0601141}}].

\bibitem{Uehara:2010xi}
S.~Uehara, \emph{{Dilaton coupling revisited}},
  \href{https://doi.org/10.1143/PTP.124.581}{\emph{Prog. Theor. Phys.}
  {\bfseries 124} (2010) 581}
  [\href{https://arxiv.org/abs/1007.5156}{{\ttfamily 1007.5156}}].

\bibitem{Meissner:2022lso}
K.~A. Meissner and H.~Nicolai, \emph{{Fundamental membranes and the string
  dilaton}}, \href{https://doi.org/10.1007/JHEP09(2022)219}{\emph{JHEP}
  {\bfseries 09} (2022) 219}
  [\href{https://arxiv.org/abs/2208.05822}{{\ttfamily 2208.05822}}].

\bibitem{Fradkin:1985ys}
E.~S. Fradkin and A.~A. Tseytlin, \emph{{Quantum String Theory Effective
  Action}}, \href{https://doi.org/10.1016/0550-3213(86)90522-5,
  10.1016/0550-3213(85)90559-0}{\emph{Nucl. Phys.} {\bfseries B261} (1985) 1}.

\bibitem{Tseytlin:1985kh}
A.~A. Tseytlin, \emph{{Covariant String Field Theory and Effective Action}},
  \href{https://doi.org/10.1016/0370-2693(86)91461-9}{\emph{Phys. Lett. B}
  {\bfseries 168} (1986) 63}.

\bibitem{Tseytlin:1988tv}
A.~A. Tseytlin, \emph{{Mobius Infinity Subtraction and Effective Action in
  $\sigma$ Model Approach to Closed String Theory}},
  \href{https://doi.org/10.1016/0370-2693(88)90421-2}{\emph{Phys. Lett. B}
  {\bfseries 208} (1988) 221}.

\bibitem{Tseytlin:1988rr}
A.~A. Tseytlin, \emph{{Sigma model approach to string theory}},
  \href{https://doi.org/10.1142/S0217751X8900056X}{\emph{Int. J. Mod. Phys.}
  {\bfseries A4} (1989) 1257}.

\bibitem{Ahmadain:2022tew}
A.~Ahmadain and A.~C. Wall, \emph{{Off-Shell Strings I: S-Matrix and Action}},
  \href{https://arxiv.org/abs/2211.08607}{{\ttfamily 2211.08607}}.

\bibitem{Duff:1987cs}
M.~J. Duff, T.~Inami, C.~N. Pope, E.~Sezgin and K.~S. Stelle,
  \emph{{Semiclassical Quantization of the Supermembrane}},
  \href{https://doi.org/10.1016/0550-3213(88)90316-1}{\emph{Nucl. Phys. B}
  {\bfseries 297} (1988) 515}.

\bibitem{Bergshoeff:1987qx}
E.~Bergshoeff, E.~Sezgin and P.~K. Townsend, \emph{{Properties of the
  Eleven-Dimensional Super Membrane Theory}},
  \href{https://doi.org/10.1016/0003-4916(88)90050-4}{\emph{Annals Phys.}
  {\bfseries 185} (1988) 330}.

\bibitem{Mezincescu:1987kj}
L.~Mezincescu, R.~I. Nepomechie and P.~van Nieuwenhuizen, \emph{{Do
  supermembranes contain massless particles?}},
  \href{https://doi.org/10.1016/0550-3213(88)90085-5}{\emph{Nucl. Phys. B}
  {\bfseries 309} (1988) 317}.

\bibitem{Forste:1999yj}
S.~Forste, \emph{{Membrany corrections to the string anti-string potential in
  M5-brane theory}},
  \href{https://doi.org/10.1088/1126-6708/1999/05/002}{\emph{JHEP} {\bfseries
  05} (1999) 002} [\href{https://arxiv.org/abs/hep-th/9902068}{{\ttfamily
  hep-th/9902068}}].

\bibitem{Drukker:2020swu}
N.~Drukker, S.~Giombi, A.~A. Tseytlin and X.~Zhou, \emph{{Defect CFT in the 6d
  (2,0) theory from M2 brane dynamics in AdS$_7 \times$S$^4$}},
  \href{https://doi.org/10.1007/JHEP07(2020)101}{\emph{JHEP} {\bfseries 07}
  (2020) 101} [\href{https://arxiv.org/abs/2004.04562}{{\ttfamily
  2004.04562}}].

\bibitem{Fradkin:1985qd}
E.~S. Fradkin and A.~A. Tseytlin, \emph{{Nonlinear Electrodynamics from
  Quantized Strings}},
  \href{https://doi.org/10.1016/0370-2693(85)90205-9}{\emph{Phys. Lett. B}
  {\bfseries 163} (1985) 123}.

\bibitem{Tseytlin:1987ww}
A.~A. Tseytlin, \emph{{Renormalization of Mobius Infinities and Partition
  Function Representation for String Theory Effective Action}},
  \href{https://doi.org/10.1016/0370-2693(88)90857-X}{\emph{Phys. Lett. B}
  {\bfseries 202} (1988) 81}.

\bibitem{Liu:1987nz}
J.~Liu and J.~Polchinski, \emph{{Renormalization of the Mobius Volume}},
  \href{https://doi.org/10.1016/0370-2693(88)91566-3}{\emph{Phys. Lett. B}
  {\bfseries 203} (1988) 39}.

\bibitem{Andreev:1988cb}
O.~D. Andreev and A.~A. Tseytlin, \emph{{Partition Function Representation for
  the Open Superstring Effective Action: Cancellation of Mobius Infinities and
  Derivative Corrections to Born-Infeld Lagrangian}},
  \href{https://doi.org/10.1016/0550-3213(88)90148-4}{\emph{Nucl. Phys. B}
  {\bfseries 311} (1988) 205}.

\bibitem{Eberhardt:2021ynh}
L.~Eberhardt and S.~Pal, \emph{{The disk partition function in string theory}},
  \href{https://doi.org/10.1007/JHEP08(2021)026}{\emph{JHEP} {\bfseries 08}
  (2021) 026} [\href{https://arxiv.org/abs/2105.08726}{{\ttfamily
  2105.08726}}].

\bibitem{Mahajan:2021nsd}
R.~Mahajan, D.~Stanford and C.~Yan, \emph{{Sphere and disk partition functions
  in Liouville and in matrix integrals}},
  \href{https://doi.org/10.1007/JHEP07(2022)132}{\emph{JHEP} {\bfseries 07}
  (2022) 132} [\href{https://arxiv.org/abs/2107.01172}{{\ttfamily
  2107.01172}}].

\bibitem{Tseytlin:2006ak}
A.~A. Tseytlin, \emph{{On sigma model RG flow, 'central charge' action and
  Perelman's entropy}},
  \href{https://doi.org/10.1103/PhysRevD.75.064024}{\emph{Phys. Rev. D}
  {\bfseries 75} (2007) 064024}
  [\href{https://arxiv.org/abs/hep-th/0612296}{{\ttfamily hep-th/0612296}}].

\bibitem{Polyakov:1987ez}
A.~M. Polyakov, \emph{Gauge fields and strings}. Taylor \& Francis, 1987.

\bibitem{Bergshoeff:1987cm}
E.~Bergshoeff, E.~Sezgin and P.~K. Townsend, \emph{{Supermembranes and
  eleven-dimensional supergravity}},
  \href{https://doi.org/10.1016/0370-2693(87)91272-X}{\emph{Phys. Lett.}
  {\bfseries B189} (1987) 75}.

\bibitem{deWit:1998yu}
B.~de~Wit, K.~Peeters, J.~Plefka and A.~Sevrin, \emph{{The M theory two-brane
  in AdS$_4\times S^7$ and AdS$_7\times S^4$}},
  \href{https://doi.org/10.1016/S0370-2693(98)01340-9}{\emph{Phys. Lett.}
  {\bfseries B443} (1998) 153}
  [\href{https://arxiv.org/abs/hep-th/9808052}{{\ttfamily hep-th/9808052}}].

\bibitem{Pasti:1998tc}
P.~Pasti, D.~P. Sorokin and M.~Tonin, \emph{{On gauge fixed superbrane actions
  in AdS superbackgrounds}},
  \href{https://doi.org/10.1016/S0370-2693(98)01597-4}{\emph{Phys. Lett. B}
  {\bfseries 447} (1999) 251}
  [\href{https://arxiv.org/abs/hep-th/9809213}{{\ttfamily hep-th/9809213}}].

\bibitem{Claus:1998fh}
P.~Claus, \emph{{Super M-brane actions in AdS(4) x S**7 and AdS(7) x S**4}},
  \href{https://doi.org/10.1103/PhysRevD.59.066003}{\emph{Phys. Rev. D}
  {\bfseries 59} (1999) 066003}
  [\href{https://arxiv.org/abs/hep-th/9809045}{{\ttfamily hep-th/9809045}}].

\bibitem{Park:2020hgt}
J.~Park and H.~Shin, \emph{{1/2-BPS membrane instantons in AdS$_4 \times$S$^7 /
  \mathbf{Z}_k$}},
  \href{https://doi.org/10.1103/PhysRevD.102.126021}{\emph{Phys. Rev. D}
  {\bfseries 102} (2020) 126021}
  [\href{https://arxiv.org/abs/2008.11380}{{\ttfamily 2008.11380}}].

\bibitem{Wu:1976ge}
T.~T. Wu and C.~N. Yang, \emph{{Dirac Monopole without Strings: Monopole
  Harmonics}}, \href{https://doi.org/10.1016/0550-3213(76)90143-7}{\emph{Nucl.
  Phys. B} {\bfseries 107} (1976) 365}.

\bibitem{Dolan:2003bj}
B.~P. Dolan, \emph{{The Spectrum of the Dirac Operator on Coset Spaces with
  Homogeneous Gauge Fields}},
  \href{https://doi.org/10.1088/1126-6708/2003/05/018}{\emph{JHEP} {\bfseries
  05} (2003) 018} [\href{https://arxiv.org/abs/hep-th/0304037}{{\ttfamily
  hep-th/0304037}}].

\bibitem{Dolan:2020sjq}
B.~P. Dolan and A.~Hunter-McCabe, \emph{{Ground State Wave Functions for the
  Quantum Hall Effect on a Sphere and the Atiyah-Singer Index Theorem}},
  \href{https://doi.org/10.1088/1751-8121/ab85e1}{\emph{J. Phys. A} {\bfseries
  53} (2020) 215306} [\href{https://arxiv.org/abs/2001.02208}{{\ttfamily
  2001.02208}}].

\bibitem{Deguchi:2005qd}
S.~Deguchi and K.~Kitsukawa, \emph{{Charge Quantization Conditions Based on the
  Atiyah-Singer Index Theorem}},
  \href{https://doi.org/10.1143/PTP.115.1137}{\emph{Prog. Theor. Phys.}
  {\bfseries 115} (2006) 1137}
  [\href{https://arxiv.org/abs/hep-th/0512063}{{\ttfamily hep-th/0512063}}].

\bibitem{Abrikosov:2002jr}
A.~A. Abrikosov, Jr., \emph{{Dirac Operator on the Riemann Sphere}},
  \href{https://arxiv.org/abs/hep-th/0212134}{{\ttfamily hep-th/0212134}}.

\bibitem{Borokhov:2002ib}
V.~Borokhov, A.~Kapustin and X.-k. Wu, \emph{{Topological Disorder Operators in
  Three-Dimensional Conformal Field Theory}},
  \href{https://doi.org/10.1088/1126-6708/2002/11/049}{\emph{JHEP} {\bfseries
  11} (2002) 049} [\href{https://arxiv.org/abs/hep-th/0206054}{{\ttfamily
  hep-th/0206054}}].

\bibitem{Pufu:2013vpa}
S.~S. Pufu, \emph{{Anomalous Dimensions of Monopole Operators in
  Three-Dimensional Quantum Electrodynamics}},
  \href{https://doi.org/10.1103/PhysRevD.89.065016}{\emph{Phys. Rev. D}
  {\bfseries 89} (2014) 065016}
  [\href{https://arxiv.org/abs/1303.6125}{{\ttfamily 1303.6125}}].

\bibitem{Dyer:2013fja}
E.~Dyer, M.~Mezei and S.~S. Pufu, \emph{{Monopole Taxonomy in Three-Dimensional
  Conformal Field Theories}},
  \href{https://arxiv.org/abs/1309.1160}{{\ttfamily 1309.1160}}.

\bibitem{Allen:1983dg}
B.~Allen, \emph{{Phase Transitions in de Sitter Space}},
  \href{https://doi.org/10.1016/0550-3213(83)90470-4}{\emph{Nucl.Phys.}
  {\bfseries B226} (1983) 228}.

\bibitem{Polyakov:1975yp}
A.~M. Polyakov and A.~A. Belavin, \emph{{Metastable States of Two-Dimensional
  Isotropic Ferromagnets}}, {\emph{JETP Lett.} {\bfseries 22} (1975) 245}.

\bibitem{DAdda:1978vbw}
A.~D'Adda, M.~Luscher and P.~Di~Vecchia, \emph{{A 1/N Expandable Series of
  Nonlinear Sigma Models with Instantons}},
  \href{https://doi.org/10.1016/0550-3213(78)90432-7}{\emph{Nucl. Phys. B}
  {\bfseries 146} (1978) 63}.

\bibitem{Din:1979zv}
A.~M. Din, P.~Di~Vecchia and W.~J. Zakrzewski, \emph{{Quantum Fluctuations in
  One Instanton Sector of the {CP}$^{n-1}$ Model}},
  \href{https://doi.org/10.1016/0550-3213(79)90280-3}{\emph{Nucl. Phys. B}
  {\bfseries 155} (1979) 447}.

\bibitem{Narain:1983in}
K.~S. Narain, \emph{{Instantons and Condensates in Supersymmetric CP$^{N-1}$
  Models}}, \href{https://doi.org/10.1016/0550-3213(84)90390-0}{\emph{Nucl.
  Phys. B} {\bfseries 243} (1984) 131}.

\bibitem{Perelomov:1987va}
A.~M. Perelomov, \emph{{Chiral Models: Geometrical Aspects}},
  \href{https://doi.org/10.1016/0370-1573(87)90044-5}{\emph{Phys. Rept.}
  {\bfseries 146} (1987) 135}.

\bibitem{Drukker:2000ep}
N.~Drukker, D.~J. Gross and A.~A. Tseytlin, \emph{{Green-Schwarz string in
  AdS(5) x S5: Semiclassical partition function}},
  \href{https://doi.org/10.1088/1126-6708/2000/04/021}{\emph{JHEP} {\bfseries
  04} (2000) 021} [\href{https://arxiv.org/abs/hep-th/0001204}{{\ttfamily
  hep-th/0001204}}].

\bibitem{Forini:2015mca}
V.~Forini, V.~G.~M. Puletti, L.~Griguolo, D.~Seminara and E.~Vescovi,
  \emph{{Remarks on the Geometrical Properties of Semiclassically Quantized
  Strings}}, \href{https://doi.org/10.1088/1751-8113/48/47/475401}{\emph{J.
  Phys. A} {\bfseries 48} (2015) 475401}
  [\href{https://arxiv.org/abs/1507.01883}{{\ttfamily 1507.01883}}].

\bibitem{Novikov:1984ac}
V.~A. Novikov, M.~A. Shifman, A.~I. Vainshtein and V.~I. Zakharov,
  \emph{{Two-Dimensional Sigma Models: Modeling Nonperturbative Effects of
  Quantum Chromodynamics}},
  \href{https://doi.org/10.1016/0370-1573(84)90021-8}{\emph{Phys. Rept.}
  {\bfseries 116} (1984) 103}.

\bibitem{Bergman:2009zh}
O.~Bergman and S.~Hirano, \emph{{Anomalous Radius Shift in Ad$S^4$/CFT(3)}},
  \href{https://doi.org/10.1088/1126-6708/2009/07/016}{\emph{JHEP} {\bfseries
  07} (2009) 016} [\href{https://arxiv.org/abs/0902.1743}{{\ttfamily
  0902.1743}}].

\bibitem{Calvo:2012du}
F.~Calvo and M.~Mari\~no, \emph{{Membrane Instantons from a Semiclassical
  TBA}}, \href{https://doi.org/10.1007/JHEP05(2013)006}{\emph{JHEP} {\bfseries
  05} (2013) 006} [\href{https://arxiv.org/abs/1212.5118}{{\ttfamily
  1212.5118}}].

\bibitem{Hatsuda:2013yua}
Y.~Hatsuda, M.~Honda, S.~Moriyama and K.~Okuyama, \emph{{ABJM Wilson Loops in
  Arbitrary Representations}},
  \href{https://doi.org/10.1007/JHEP10(2013)168}{\emph{JHEP} {\bfseries 10}
  (2013) 168} [\href{https://arxiv.org/abs/1306.4297}{{\ttfamily 1306.4297}}].

\bibitem{Okuyama:2016deu}
K.~Okuyama, \emph{{Instanton Corrections of 1/6 BPS Wilson Loops in ABJM
  Theory}}, \href{https://doi.org/10.1007/JHEP09(2016)125}{\emph{JHEP}
  {\bfseries 09} (2016) 125}
  [\href{https://arxiv.org/abs/1607.06157}{{\ttfamily 1607.06157}}].

\bibitem{Aganagic:2001nx}
M.~Aganagic, A.~Klemm and C.~Vafa, \emph{{Disk Instantons, Mirror Symmetry and
  the Duality Web}}, \href{https://doi.org/10.1515/zna-2002-1-201}{\emph{Z.
  Naturforsch. A} {\bfseries 57} (2002) 1}
  [\href{https://arxiv.org/abs/hep-th/0105045}{{\ttfamily hep-th/0105045}}].

\bibitem{Hatsuda:2012hm}
Y.~Hatsuda, S.~Moriyama and K.~Okuyama, \emph{{Exact Results on the ABJM Fermi
  Gas}}, \href{https://doi.org/10.1007/JHEP10(2012)020}{\emph{JHEP} {\bfseries
  10} (2012) 020} [\href{https://arxiv.org/abs/1207.4283}{{\ttfamily
  1207.4283}}].

\bibitem{Hatsuda:2013oxa}
Y.~Hatsuda, M.~Mari\~no, S.~Moriyama and K.~Okuyama, \emph{{Non-Perturbative
  Effects and the Refined Topological String}},
  \href{https://doi.org/10.1007/JHEP09(2014)168}{\emph{JHEP} {\bfseries 09}
  (2014) 168} [\href{https://arxiv.org/abs/1306.1734}{{\ttfamily 1306.1734}}].

\bibitem{Bagger:2012jb}
J.~Bagger, N.~Lambert, S.~Mukhi and C.~Papageorgakis, \emph{{Multiple Membranes
  in M-theory}},
  \href{https://doi.org/10.1016/j.physrep.2013.01.006}{\emph{Phys. Rept.}
  {\bfseries 527} (2013) 1} [\href{https://arxiv.org/abs/1203.3546}{{\ttfamily
  1203.3546}}].

\bibitem{Halyo:1998pn}
E.~Halyo, \emph{{Supergravity on AdS(5/4) $\times$ Hopf Fibrations and
  Conformal Field Theories}},
  \href{https://doi.org/10.1016/S0217-7323(00)00038-4}{\emph{Mod. Phys. Lett.
  A} {\bfseries 15} (2000) 397}
  [\href{https://arxiv.org/abs/hep-th/9803193}{{\ttfamily hep-th/9803193}}].

\bibitem{Gustavsson:2009pm}
A.~Gustavsson and S.-J. Rey, \emph{{Enhanced ${\mathcal{N}}\!=8$ Supersymmetry
  of ABJM Theory on $ R^8$ and $ R^8$/$Z_2$}},
  \href{https://arxiv.org/abs/0906.3568}{{\ttfamily 0906.3568}}.

\bibitem{Putrov:2012zi}
P.~Putrov and M.~Yamazaki, \emph{{Exact ABJM Partition Function from TBA}},
  \href{https://doi.org/10.1142/S0217732312502008}{\emph{Mod. Phys. Lett. A}
  {\bfseries 27} (2012) 1250200}
  [\href{https://arxiv.org/abs/1207.5066}{{\ttfamily 1207.5066}}].

\end{thebibliography}\endgroup
\bibliographystyle{JHEP-v2.9}
\end{document}